\documentclass[conference]{IEEEtran}

\usepackage{fancyhdr}
\usepackage{hyperref}

\usepackage{graphicx, amsmath,nccmath}
\usepackage{enumitem}
\usepackage{algpseudocode, algorithm, subfiles}
\usepackage{booktabs}

\usepackage{subfig}
\usepackage[skip=5pt,font=footnotesize]{caption}

\usepackage{adjustbox} 

\usepackage{url}

\newcommand{\ignore}[1]{}

\title{Rendering Elimination: Early Discard of Redundant Tiles in the Graphics Pipeline}
\begin{document}
%



%
\author{\IEEEauthorblockN{Mart\'{i} Anglada\IEEEauthorrefmark{1},
Enrique de Lucas\IEEEauthorrefmark{1},
Joan-Manuel Parcerisa\IEEEauthorrefmark{1},
Juan L. Arag\'{o}n\IEEEauthorrefmark{2},
Antonio Gonz\'{a}lez\IEEEauthorrefmark{1},
Pedro Marcuello\IEEEauthorrefmark{3}}
\IEEEauthorblockA{\IEEEauthorrefmark{1}Universitat Polit\`{e}cnica de Catalunya. \{manglada, edelucas, jmanel, antonio\}@ac.upc.edu}
\IEEEauthorblockA{\IEEEauthorrefmark{2}Universidad de Murcia. jlaragon@ditec.um.es}
\IEEEauthorblockA{\IEEEauthorrefmark{3}Broadcom Corporation. pedro.marcuello@broadcom.com}}


\maketitle
\pagestyle{plain}

\begin{abstract}
GPUs are one of the most energy-consuming components for real-time rendering applications, since a large number of fragment shading computations and memory accesses are involved. Main memory bandwidth is especially taxing battery-operated devices such as smartphones. Tile-Based Rendering GPUs divide the screen space into multiple tiles that are independently rendered in on-chip buffers, thus reducing memory bandwidth and energy consumption. We have observed that, in many animated graphics workloads, a large number of screen tiles have the same color across adjacent frames. In this paper, we propose \textit{Rendering Elimination} (RE), a novel micro-architectural technique that accurately determines if a tile will be identical to the same tile in the preceding frame before rasterization by means of comparing signatures. Since RE identifies redundant tiles early in the graphics pipeline, it completely avoids the computation and memory accesses of the most power consuming stages of the pipeline, which substantially reduces the execution time and the energy consumption of the GPU. For widely used Android applications, we show that RE achieves an average speedup of 1.74x and energy reduction of 43\% for the GPU/Memory system, surpassing by far the benefits of Transaction Elimination, a state-of-the-art memory bandwidth reduction technique available in some commercial Tile-Based Rendering GPUs.

\end{abstract}

\section{Introduction}\label{sec:intro}
\graphicspath{{Introduction/}}


Graphics applications for smartphones and tablets have become ubiquitous platforms for entertainment, with more than 2 billion users worldwide and more than a 40\% share of the overall games market~\cite{newzooUsers}. The portable nature of such devices drives engagement to games with simple gameplay that can be played in short bursts, such as puzzle, strategy or casual games, genres that represent the greatest number of downloads and played time~\cite{esa2014, medium2016, google2017}. While games of those characteristics usually do not involve complex scenes and cutting-edge effects, rendering their scenes still requires a substantial amount of power, a limited resource in battery-operated devices. Consequently, reducing the energy consumption of the GPU is a major concern of hardware and software designers~\cite{Clarberg:2013:SDS:2461912.2462022, wang2010kernel, deLucas:2015:UPR:2830772.2830783, chu2011energy,nam2007low}.

\begin{figure}[ht]
\centering
   \includegraphics[width=1\linewidth]{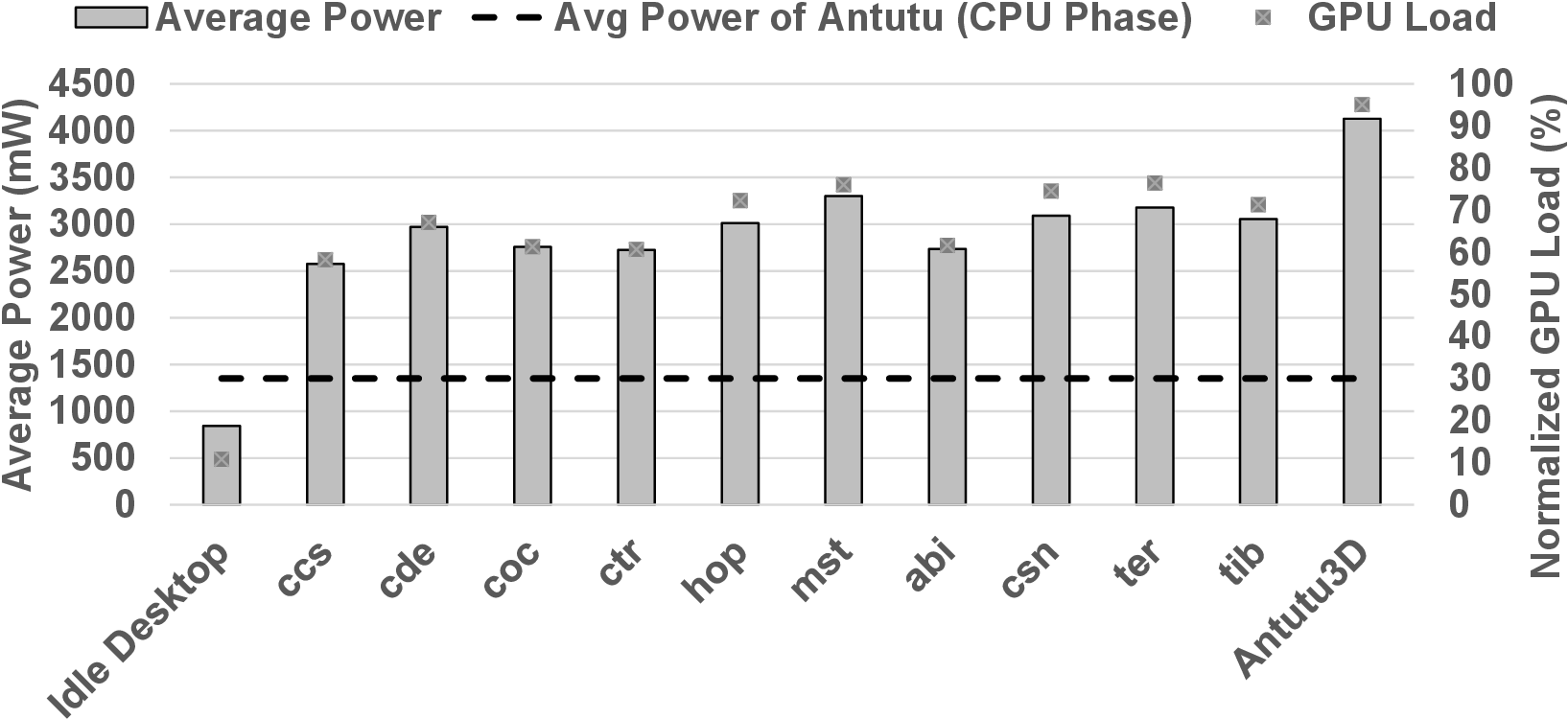}
\caption[]{Overall average power consumption. GPU load is normalized by weighting it by the ratio between operating and maximum GPU frequency. Data obtained using Trepn Profiler~\cite{trepn} for a Snapdragon 636 with connections disabled and minimum screen brightness during a few minutes.}
 \label{fig:absolutepower} 
  \vspace{-1em} 
\end{figure}

Figure~\ref{fig:absolutepower} shows the average power consumption and GPU load for the Android desktop (without animations), for several commercial Android games and the Antutu benchmark~\cite{antutu}, divided into the CPU phase and the GPU phase (\textit{Antutu3D}). As it can be seen, applications with simple scenes such as CandyCrush (\textit{ccs}) require a substantial amount of power and GPU load, comparable to an application designed to stress the GPU. Note that they also drive much more power than the Android desktop, an application that lets the GPU mostly idle, while consuming twice as much as an application that only stresses the CPU. These  experimental results confirm the popular claim that, in graphics applications, the GPU and its communication with main memory (loading textures and storing colors, among other tasks) are the greatest contributors to energy consumption~\cite{Akenine-Moller:2003:GMH:882262.882348,carroll2010analysis,nystad2012adaptive}.

A state-of-the-art pipeline design employed to reduce bandwidth in mobile GPUs is \textit{Tile-Based Rendering} (TBR). In TBR, a frame space is divided into a grid of tiles that are independently rendered, 
which allows to do a variety of computations leveraging small, fast, local on-chip memory instead of using main memory. The graphics pipeline in a TBR GPU is divided into two decoupled pipelines: the Geometry Pipeline receives vertices and generates, after a set of transformations, output primitives (triangles) that are sorted into tile bins and stored into the main memory Parameter Buffer; and the Raster Pipeline which traverses the tiles one at a time, fetching each tile's primitives, rasterizing each primitive into fragments, and shading each fragment to obtain a final pixel color.

\begin{figure}[ht]
\centering
   \includegraphics[width=1\linewidth]{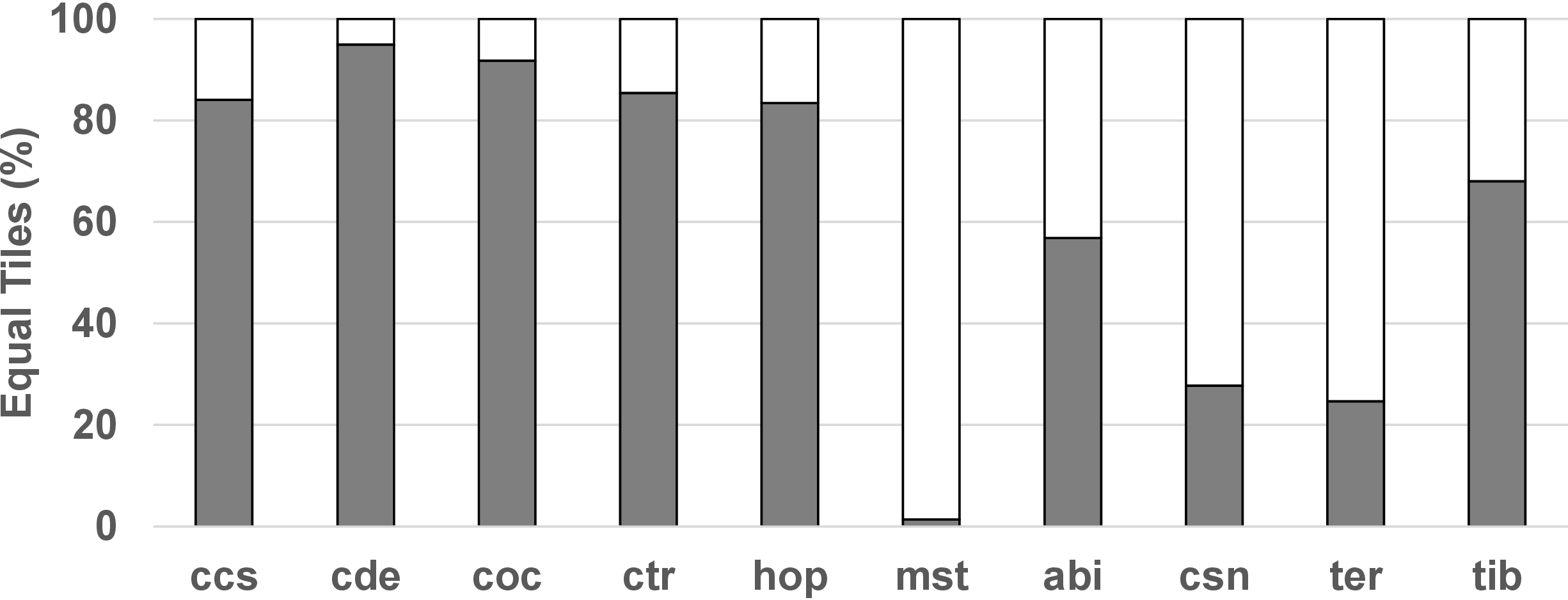}
\caption[]{Percentage of tiles producing the same color as the preceding frame across 50 consecutive frames (experimental details in Section \ref{sec:methodology}).}
 \label{fig:equaltiles_i} 
   \vspace{-0.25em}
\end{figure}



A main purpose of the GPU is to render sequences of images. In order to produce fluid animations, consecutive frames tend to be similar, i.e., it is usual to find regions in a frame with the same color as in the preceding frame, which implies that a significant amount of computations are redundant. Figure~\ref{fig:equaltiles_i} illustrates this phenomenon, known as Frame-to-Frame coherence \cite{Hubschman:1982}, by plotting the average percentage of equal tiles between two consecutive frames for a set of commercial Android games. In games with moderate camera movements (\textit{ccs} to \textit{hop}), over 90\% of tiles produce the same color as in the preceding frame. This feature can also be found, albeit less frequently, in games where the scene is in continuous motion (\textit{mst} to \textit{tib}).

Several previous works attempted to exploit frame-to-frame coherence in order to improve energy efficiency. Transaction Elimination~\cite{TE} (TE) compares a signature of the colors generated after rendering a tile with the signature of the same tile in the preceding frame. If they are equal, the color update to main memory is avoided. Arnau et al. \cite{Arnau:2014:ERF:2665671.2665748} proposed a task-level Fragment Memoization scheme that computes, for each fragment, a signature of all its shader inputs and caches it, along with the output color, in a LUT. Subsequent fragments form their signatures and check them against the signatures of the memoized fragments. In case of a hit, the shader's computation and associated texture accesses are avoided and the cached color is used instead. Because most redundancy resides between consecutive frames, the huge reuse distance makes impractical to store a frame's worth of signatures and output values. To help reduce the reuse distance, it builds on top of PFR~\cite{Arnau:2013:PFR:2523721.2523736}, an architecture that renders two consecutive frames in parallel and keeps tiles synchronized. But PFR cuts in half the redundancy detection potential: even frames reuse values cached by the previous (odd) frame, but odd frames cannot because their previous-frame values are already evicted from the LUT by the time they are rendered.

We make the observation that in a TBR GPU, primitives do not need to be discretized into fragments to know that the final result will be the same as in the preceding frame. Instead, by managing redundancy at a tile level, redundant tiles may be discovered much earlier than at a fragment level and bypass the whole Raster Pipeline, not just the Fragment Shader stage. Note that the Raster Pipeline computes the pixel colors using as inputs a set of primitives' attributes generated by the Primitive Assembly stage of the Geometry Pipeline plus a set of scene constants, so it knows all the input data required to render a tile when it starts processing it.

Based on the above observation, we propose \textit{Rendering Elimination} (RE), a novel technique that employs the input data of a tile to anticipate if all of its pixels will have the same color as in the preceding frame, and to bypass the complete rendering of the tile. Since an entire frame of these input sets must be stored on-chip, they are compared by means of a signature. In parallel with the sorting of a primitive into tiles, RE computes on-the-fly the signatures of the overlapped tiles and stores them in a local fixed-size on-chip buffer. Then, after the Geometry Pipeline has processed the frame, tiles are dispatched to the Raster Pipeline. For each tile, RE compares its current and preceding frame signatures and, if they match, all the rendering process is bypassed and the colors in the Frame Buffer are reused. Otherwise, the tile is rendered as usual.

\begin{figure}[ht]
\centering
   \includegraphics[width=1\linewidth]{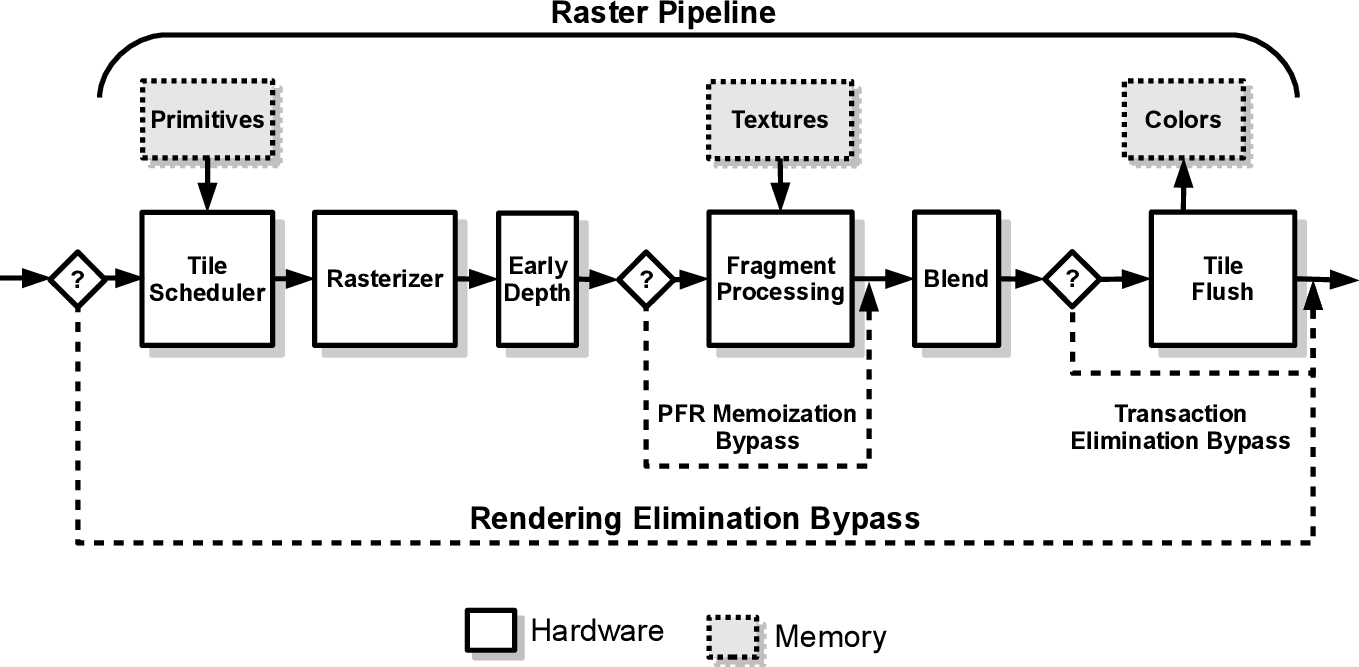}
\caption[]{Raster Pipeline stages saved by using Transaction Elimination, Fragment Memoization or Rendering Elimination.}
\label{fig:f2fraster}
\vspace{-1em}
\end{figure}

By working at a much coarser grain than Fragment Memoization \cite{Arnau:2014:ERF:2665671.2665748}, RE can store on-chip all the frame signatures and detect all the available tile redundancy instead of just that of the even frames, which more than compensates for the marginal undetected redundancy at sub-tile level (our results show that RE almost doubles the amount of redundancy discovered). In addition, RE does not need to store output results because tile colors are reused from the Frame Buffer, thus saving storage and bandwidth. Besides this, while TE and Fragment Memoization each skip just a single stage of the Raster Pipeline (as depicted in Figure \ref{fig:f2fraster}), RE completely skips all the Raster Pipeline stages. Considering that almost 75\% of the total GPU memory accesses (textures, colors and primitives) are generated by these stages, our approach is able to greatly reduce memory bandwidth and energy consumption.

The main contributions of this paper are: (1) The observation that frame-to-frame redundancy can be discovered in a TBR GPU at the tile level much earlier in the pipeline than previous techniques do. (2) A detailed proposal of Rendering Elimination, a technique for early discarding of redundant tiles, clearly showing how RE may be seamlessly integrated into the Graphics Pipeline with minimal hardware and performance overheads. (3) An experimental evaluation of RE that shows an average speedup of 1.74x and 43\% energy reduction over a conventional mobile GPU, and substantial improvements over previous works.

The remainder of this paper is organized as follows. Section \ref{sec:background} overviews the fundamentals of our baseline TBR graphics pipeline. Section \ref{sec:re} details the proposed RE approach. Section \ref{sec:methodology} describes the experimental framework. Section \ref{sec:results} shows energy and performance results. Section \ref{sec:rw} reviews related work and Section \ref{sec:conclusions} contains the conclusions.

\section{Tile-Based Rendering Baseline}\label{sec:background}
\graphicspath{{Background/}}

\begin{figure}[h]
\centering
   \includegraphics[width=\linewidth]{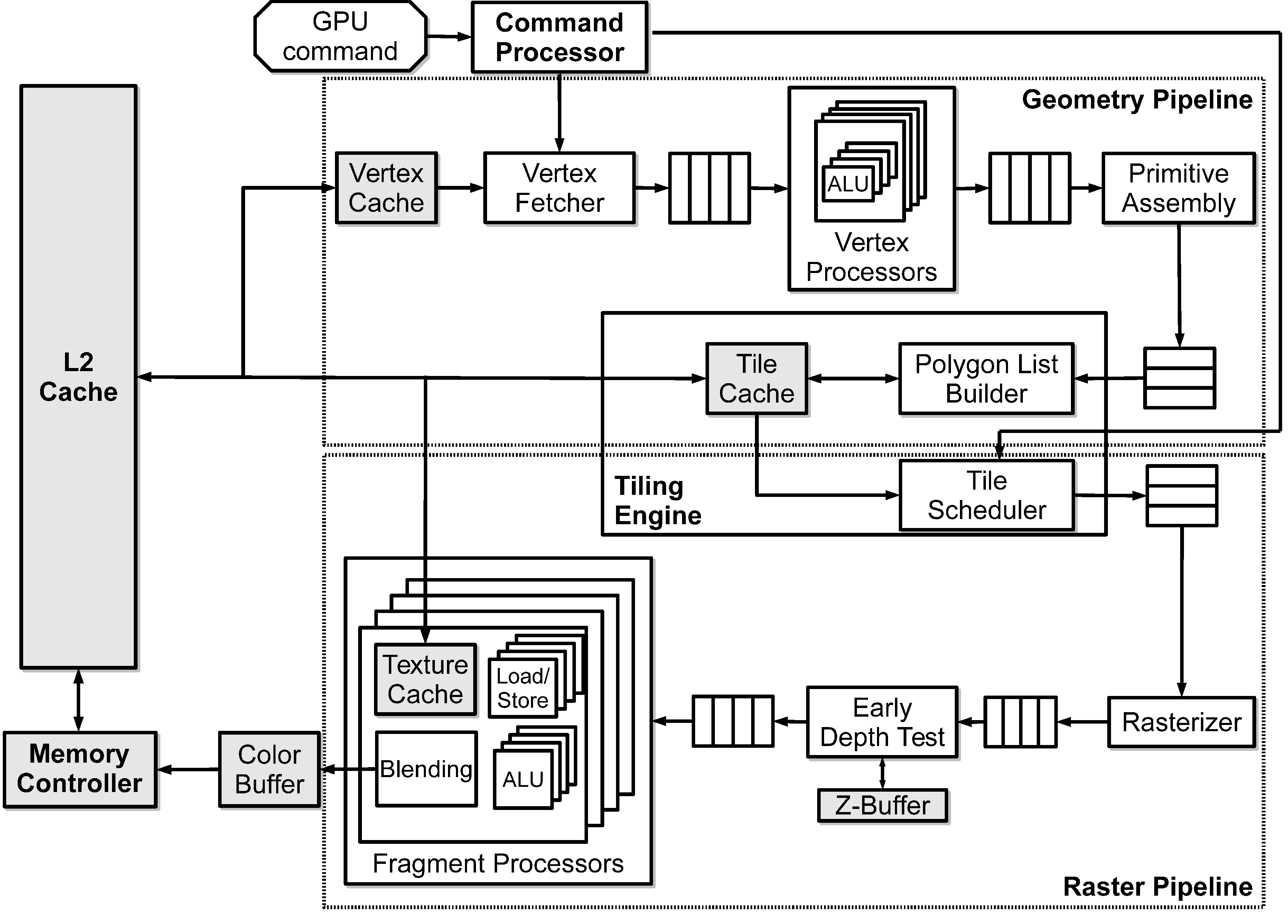}
   \caption{Assumed baseline architecture.}
   \label{fig:TBR_b} 
\vspace{-0.5em}
\end{figure}

Figure \ref{fig:TBR_b} shows the baseline architecture used in this paper, which resembles an ARM Mali-450 GPU~\cite{mali450}. While it is only a single point in the design spectrum of GPUs, it serves perfectly to demonstrate the huge benefits of our technique for a broad class of GPU architectures with the only requirement to follow an OpenGL compliant TBR organization. This architecture is a hardware implementation of the Graphics Pipeline, a conceptual model that describes the stages through which data should be processed in order to render a scene. The application communicates with the GPU using commands, which are used to configure the pipeline state (shader code, constants --``uniforms''--, textures, ...) and to trigger execution via \textit{drawcalls}, a stream of vertices to be processed with the current state. The pipeline state is held constant during a drawcall invocation but may be altered between invocations.

The Command Processor parses drawcalls and determines the format used by the application to submit vertices to the pipeline. Next, the Vertex Fetcher creates an input stream of vertices by reading information with the established format. The per-vertex read information is known as \textit{Vertex Attributes}, and consists of sets of data that specify vertices, such as 3D space coordinates or color. The vertex stream is then \textit{shaded}: the attributes of each vertex are transformed using Vertex Processors that execute programs set by the application. These programs are called shaders, and are shared among all vertices of a drawcall. The shaded vertices are grouped into triangles or other primitives in the final stage of the Geometry Pipeline, known as Primitive Assembly, where some of the non-visible ones are discarded applying clipping and culling techniques.

The primitives resulting from the geometry process are sent to the Tiling Engine, where the Polygon List Builder stores primitives' attributes in a region of memory known as Parameter Buffer. The attributes are stored in a format that exploits locality and enhances performance on the Raster Pipeline. The Polygon List Builder also determines in which tiles each primitive resides. After all the geometry has been sorted into tiles and saved in the Parameter Buffer, the tiles are processed in sequence. The Tile Scheduler is responsible of fetching the primitives' data for a given tile and dispatching it to the Raster Pipeline.

The Raster Pipeline starts by rasterizing primitives. The primitives are discretized into fragments: pixel-sized elements described by interpolated information from vertex attributes. The Early Depth Test is used to discard fragments that would be occluded by previously processed fragments. The fragments that pass the Early Depth Test are sent to the Fragment Processors, which execute application-defined shaders to compute the color for every fragment. The output color computed in the Fragment Shaders for a given pixel is merged with the previously computed colors using the Blending unit, and the resulting color is written into the local on-chip Color Buffer. When the Raster Pipeline has processed all of the primitives of a tile, the contents of the Color Buffer are flushed into the Frame Buffer in system memory and the Raster Pipeline begins processing the next tile.

\section{Rendering Elimination}\label{sec:re}
\subsection{Overview}
\graphicspath{{RenderingElimination/}}

This paper proposes Rendering Elimination, a novel micro-architectural technique that accurately determines if a tile is redundant, i.e., if all of its pixels will have the same color as they had in the previous frame. Whenever a tile is detected as redundant, its Raster Pipeline execution is completely bypassed and the color from the previous frame is reused.

\begin{figure}[ht]
\centering
   \includegraphics[width=\linewidth]{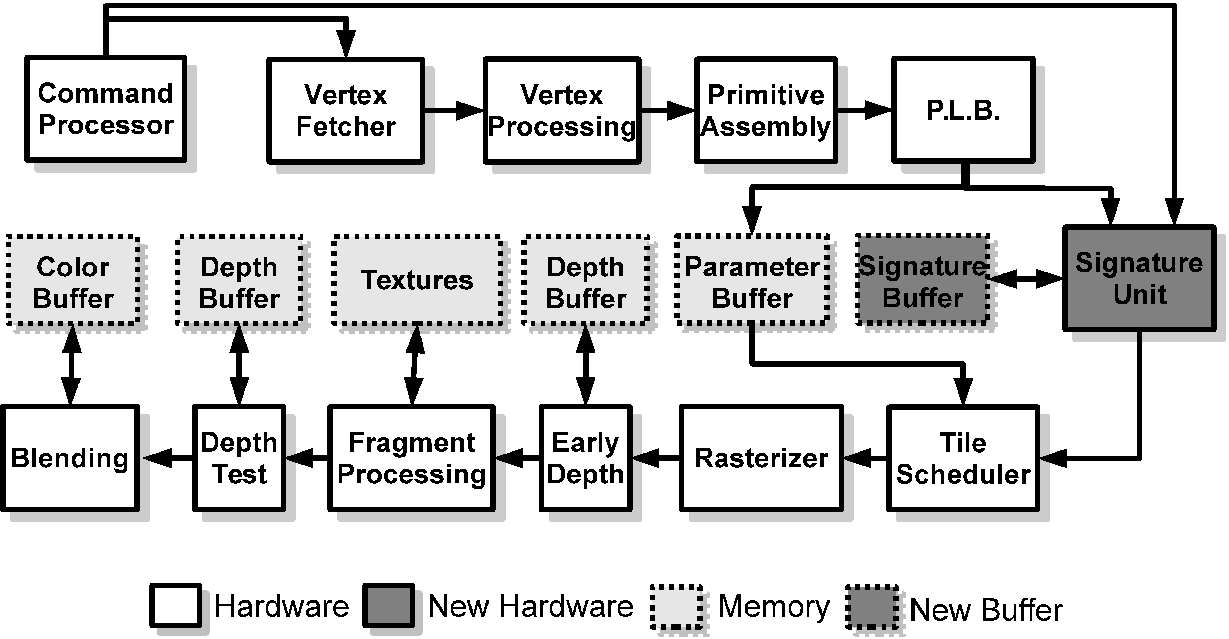}
   \caption{Graphics Pipeline including Rendering Elimination.}
   \label{fig:new_pipe_re} 
\end{figure}

The Raster Pipeline takes as inputs the scene constants and the attributes of all the primitives that overlap a tile, and produces a color for each pixel belonging to that tile. In order to determine in advance redundancy for a tile, we compare its inputs for the current frame against the inputs for the previous frame: if the two input sets match, the outputs will also be equal. Because of the large volume of these sets, storing them in main memory would be extremely inefficient, even with the support of a cache, because the reuse distance between them is an entire frame. Instead, we use a more efficient approach based on computing a signature for the inputs of the tile and storing it in a local buffer. This buffer, that we call \textit{Signature Buffer}, contains the signatures of all the tiles of the previous and current frames. Figure \ref{fig:new_pipe_re} depicts the Graphics Pipeline flow with the added Signature Buffer. 

The Signature Unit computes the signatures employing the primitives that the Polygon List Builder produces and inserts them into the Signature Buffer. At the same time, the Polygon List Builder fills the Parameter Buffer with the data of such primitives, including identifiers of the tiles that contain them. After the geometry of the frame has been processed, the Signature Buffer holds signatures for the inputs of all the tiles. Hereafter, whenever a tile is scheduled in the Raster Pipeline, its Signature Buffer entry is checked: if the current frame signature matches that of the previous frame, the Raster Pipeline execution is skipped and the Frame Buffer locations for that tile are not updated. Otherwise, the Raster Pipeline is executed normally.

\subsection{Implementation Requirements}

The signature of a tile is computed by hashing a list of all the inputs of a tile: this includes the vertex attributes and scene constants associated to all the primitives that overlap the tile. Such inputs are produced either by the Command Processor when setting scene constants for a drawcall or by the Polygon List Builder when sorting primitives and storing their vertex attributes into the Parameter Buffer. The stream of primitives produced by the Geometry Pipeline, however, is generated in the order that the GPU received the drawcalls, which is generally not the order in which they appear in the screen. In fact, any primitive from the stream could overlap any number of tiles. This causes that the complete list of inputs for a tile is not known until all the geometry of the scene has been processed. A straightforward implementation that starts computing the signatures when the Geometry Pipeline has processed the whole frame would not be practical. Since vertex attributes are stored in the Parameter Buffer (residing in off-chip memory), retrieving them in order to compute a signature for the tile would require significant time and energy overheads and delaying the  execution of the Raster Pipeline.

To be effective, our technique computes the signatures for the current frame in an \textit{incremental} approach. Whenever a primitive is sorted, the temporary signatures for each tile that it overlaps are read. The new signature for each tile is constructed by combining the temporary signature with either the scene constants or the attributes of the vertices of the current primitive and, afterwards, it is rewritten in the appropriate Signature Buffer entry. This on-the-fly signature computation is overlapped with other Geometry Pipeline stages, resulting in minimal overheads in execution time.

The signature function employed by Rendering Elimination is CRC32 \cite{peterson1961cyclic}. While a plethora of other mechanisms exist, CRC32 outperforms well-known hashing approaches such as XOR-based schemes, as we will show in Section \ref{sec:results}. We have not observed a single instance of hashing collisions in our benchmarks when using CRC32. Moreover, as a widely-used error detection code, CRC has been extensively researched in the literature and efficient techniques have been developed \cite{sun2013high} that allow for an incremental and parallel CRC computation based on Look-up Tables (LUTs), as outlined below.

\subsection{Incremental CRC32 Computation}\label{sec:incremental}
As proven in \cite{sun2013high}, the CRC of a message can be computed even if its length is not known a priori by breaking it down into several submessages and computing the CRC of those submessages independently. Given a message $A$, composed by concatenating submessages $A_{1}...A{n}$, of lenghts $b_{1}...b_{n}$ bits, the CRC of $A$ can be computed as:

\begin{algorithm}[h]
\begin{algorithmic}
 \State $CRC_{A} = 0$
 \For{submessage $A_{i}$ in $A$}
 \State $b = length(A_{i})$
 \State $CRC_{A_{i}} = ComputeCRC (A_{i})$
 \State $CRC_{Temporary} = ComputeCRC (CRC_{A} << b)$
 \State $CRC_{A} = CRC_{A_{i}} \oplus CRC_{Temporary}$ 
 \EndFor
 \caption{Incremental CRC Computation}
 \label{alg:incremental}
 \end{algorithmic}
\end{algorithm}

That is, the CRC of the first submessage $A_{1}$ is computed. When the length $b$ of the following submessage $A_{2}$ is known, we can compute the CRC of the two submessages (a bit string formed by concatenating $A_{1}$ and $A_{2}$) by computing the CRC of $A_{2}$, left-shifting the CRC of $A_{1}$ by $b$ bits, computing the CRC of this shifted message, and combining both CRCs via an XOR function. By means of this procedure, CRCs of partial messages of increasing length are computed: first, the CRC of $A_{1}$, then the CRC of the concatenation of $A_{1}$ and $A_{2}$, then the CRC of the concatenation of $A_{1}$, $A_{2}$ and $A_{3}$, and so on, until the last submessage $A_{n}$ is reached and, therefore, the CRC of the concatenation of the submessages corresponds to the CRC of the original message.

\subsection{Table-based CRC32 Computation} \label{sec:parallel}
Each iteration in Algorithm \ref{alg:incremental} would require several cycles if the CRC computation was implemented using the basic Shift Register mechanism \cite{massey1969shift}. A faster alternative is to use a Look-up Table (LUT) loaded with precomputed CRC values for all possible inputs. However, this approach is unfeasible in terms of storage requirements, since a message of length $n$ requires a LUT of $2^{n}$ entries. As shown in \cite{sun2013high}, a message $B$ of $n$ bits, being $n$ multiple of 8, can be broken into $k$ 1-byte blocks $B_{1}... B_{k}$ ($n=8 \times k$) and use a small LUT to efficiently compute the CRC of each block. 

Each LUT takes as input a block $B_{i}$ and computes the CRC32 of a message corresponding to left-shifting $B_{i}$ by $k-i$ bytes. Namely, the first LUT computes the CRC32 of a message consisting of block $B_{1}$ followed by $k-1$ bytes of zeros, the second LUT comptues the CRC32 of a message consisting of block $B_{2}$ followed by $k-2$ bytes of zeros and the $k^{th}$ LUT computes the CRC32 of a message consisting of block $B_{k}$. The results of the $k$ LUTs are then combined into one unique CRC via an XOR function. Note that this Parallel CRC Computation is effectively equivalent to unrolling the loop in Algorithm \ref{alg:incremental}.

Since each LUT has $2^{8}$ entries and each entry contains a precomputed CRC32 value, the size of each LUT is 1 KB and, consequently, computing the CRC32 of a message of length $n$ bits has a storage cost of $k$ KB. 

\subsection{Tile Inputs Bitstream Structure}
\graphicspath{{RenderingElimination/}}
RE determines if the colors of two tiles are going to be the same by comparing the signature of their inputs. The inputs of a tile are the vertex attributes of the primitives that overlap it and the set of scene constants associated to those primitives. In order to render primitives, the GPU receives a series of commands that define the state of the pipeline (shaders, textures, constants) and drawcalls, which contain a stream of vertices to be processed with the defined state.

Each drawcall can generate any number of primitives and each primitive can overlap any number of tiles. Therefore, the input of a tile consists of a sequence of blocks, one for every drawcall that contains the primitives that overlap this tile. Each block is, in turn, composed of several subblocks: a first subblock corresponding to the constants defined in the drawcall followed by a list of subblocks that correspond to the attributes of the primitives that overlap this tile. Since both the number of primitives overlapping a tile and the number of attributes of those primitives is not fixed, neither are the lengths of the blocks nor is the length of the subblocks.

\begin{figure}[ht]
 \centering
 \includegraphics[width=0.35\textwidth]{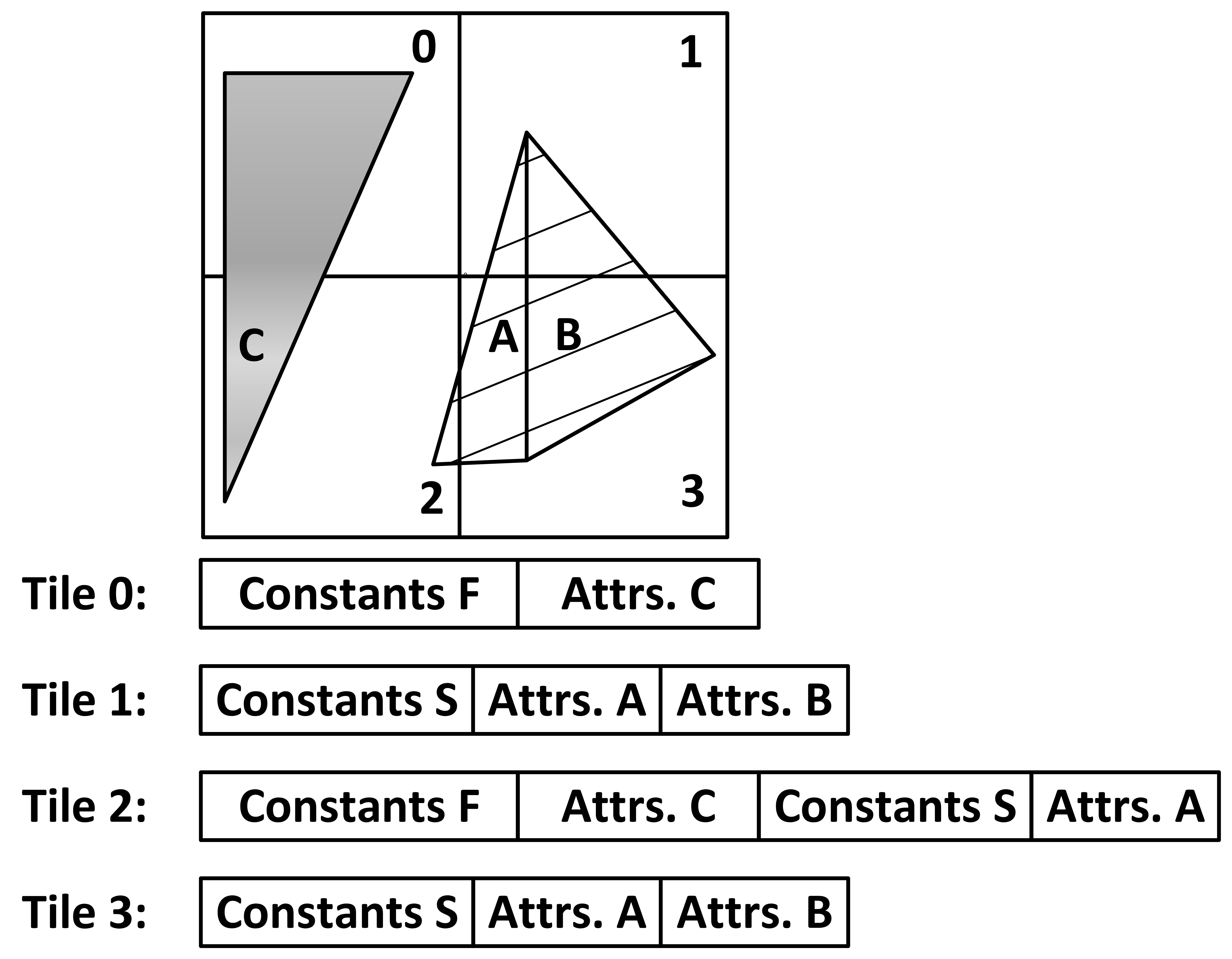}
 \caption{Example of input message for four tiles.}
 \label{fig:example}
\end{figure}

Figure \ref{fig:example} provides an example of the described tile inputs for four tiles and the primitives of two drawcalls: Drawcall F (fill) and Drawcall S (stripes). Drawcall F generates Primitive C, which overlaps Tiles 0 and 2. Therefore, the inputs of Tiles 0 and 2 contain the block of Drawcall F, composed of a set of constants and the attributes of Primitive C. Drawcall S generates two primitives, Primitives A and B. These two primitives overlap Tiles 1 and 3, so the inputs of Tiles 1 and 3 contain the block of Drawcall S, composed of a set of constants and the attributes of both primitives. Note that, while two primitives of Drawcall S overlap Tiles 1 and 3, the set of constants of the drawcall is only considered once for those tiles. Primitive A also overlaps Tile 2, so the set of constants of Drawcall S as well as the attributes of Primitive A are added to the inputs of Tile 2. 

Besides scene constants, primitives have other global associated data that affects the color of a fragment: the shader program and the textures to be used within. Rendering Elimination does not include these in the tile signature, since changes to such global data are not common. In our benchmarks, we have observed that shaders and textures remain constant for thousands of frames. Moreover, loading new shaders and textures is done through API calls (such as \textit{glShaderSource} and \textit{glTexImage2D}, for instance) and, therefore, are registered by the driver. Whenever such infrequent API calls occur, Rendering Elimination is disabled for the current frame. Besides this, RE could also be disabled during one frame periodically to guarantee Frame Buffer refreshing. RE should also be temporarily disabled by the driver for scenes that use multiple render targets: RE is specifically targeted to an important segment of less sophisticated applications that cover a large fraction of the mobile market.

\subsection{Signature Unit Architecture}
\graphicspath{{RenderingElimination/}}

The message that has to be signed for a tile consists of a sequence of blocks, containing either scene constant data or vertex attribute data. The number of blocks of a message is not known until all the geometry of the frame is processed and, therefore, Rendering Elimination uses the incremental signature computation described in Algorithm \ref{alg:incremental}.

\begin{figure}
 \centering
 \includegraphics[width=0.40\textwidth]{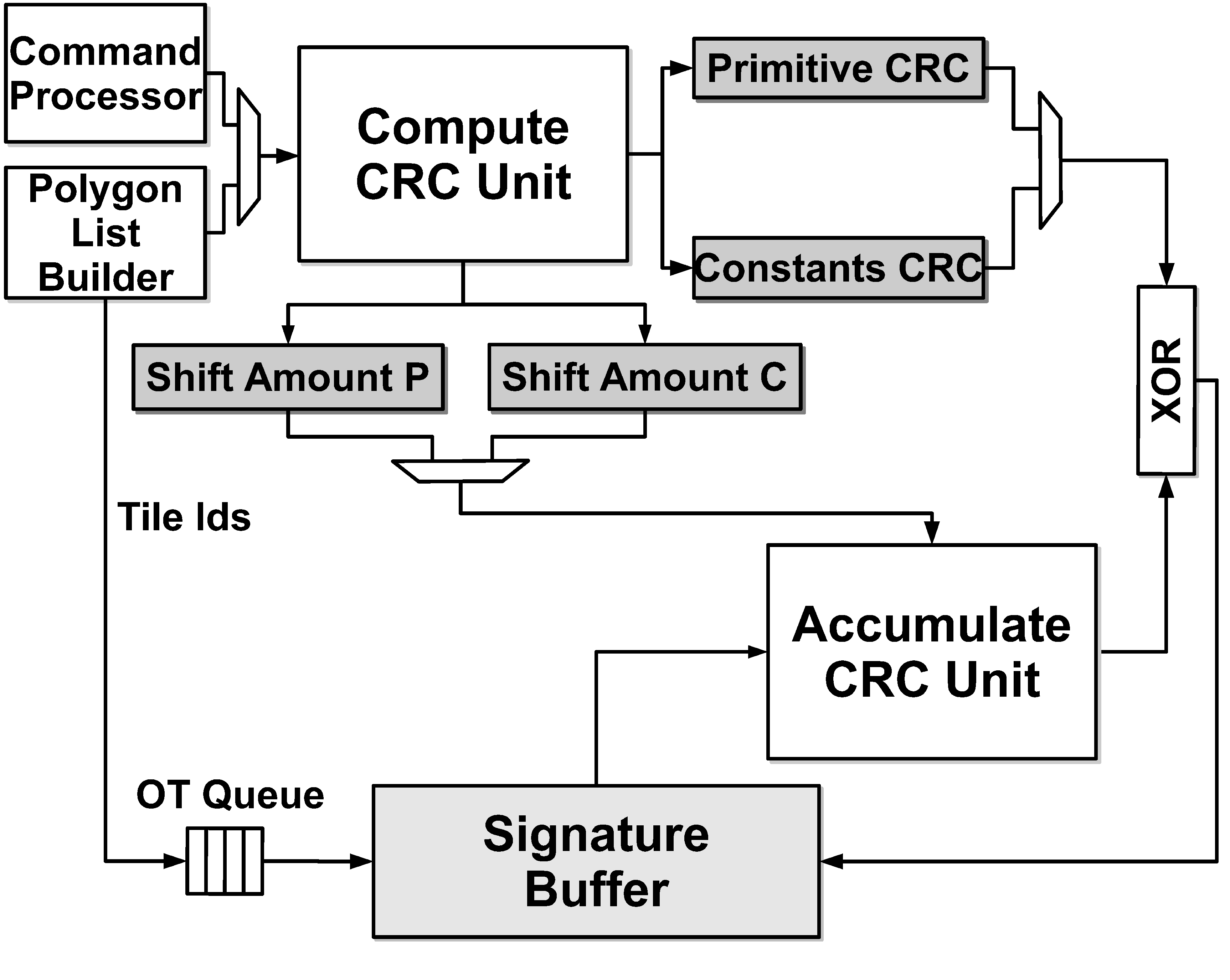}
 \caption{Signature Unit block diagram.}
 \label{fig:architecture}
 \vspace{-1em}
\end{figure}

The Signature Unit (SU), which is the piece of logic responsible for the incremental computation of the CRCs of tiles, is shown in Figure \ref{fig:architecture}. Whenever the SU receives a new data block, it computes its CRC and updates the CRC of all the tiles overlapped by the primitive associated to that block.

Let us consider first the case of vertex attributes, which are blocks sent to the SU by the Polygon List Builder. The SU computes the signature of all the vertex attributes of a primitive using the \textit{Compute CRC} unit, and the resulting CRC32 ($CRC_{A_{i}}$ as described by Algorithm \ref{alg:incremental}) is stored in the \textit{Primitive CRC} register. Since the number of attributes in a primitive is variable, the Compute CRC unit stores the length of the signed block ($b$ in Algorithm \ref{alg:incremental}) in the \textit{Shift Amount P} register. While the SU computes the CRC of a primitive, the Polygon List Builder inserts into the \textit{OT Queue} a list of identifiers of the tiles overlapped by the primitive.

After computing the signature of a primitive, the Signature Unit traverses the list of overlapped tiles and updates each tile signature by combining it with the primitive signature. It pops in sequence each entry from the head of the OT Queue and uses this tile \textit{id} to read the corresponding CRC from the Signature Buffer, which is then sent to the \textit{Accumulate CRC} unit. This unit receives as inputs the previous CRC for a tile and the length of the primitive message signed by the Compute Unit. The Accumulate CRC unit computes the CRC of the message that results by left-shifting the previous CRC as many bits as the received length. This CRC corresponds to $CRC_{Temporary}$ in Algorithm \ref{alg:incremental}. Finally, the results of the Compute and Accumulate units are bitwise xored to obtain the new CRC for the tile ($CRC_{A}$ in Algorithm \ref{alg:incremental}) and the new signature is written back to the Signature Buffer.

The Signature Unit can also receive data blocks from the Command Processor, which correspond to scene constants. The signature computation of the constants of a drawcall is done in the same form as the signature computation of the vertex attributes of a primitive: the Compute Unit generates a CRC32 and the length of the signed message and stores them in two registers: \textit{Constants CRC} and \textit{Shift Amount C}, respectively.

In order to combine the signature of the constants with the signature of the attributes, several issues need to be addressed. First, every drawcall may define its own set of constants which only affect to that drawcall. Consequently, the Constant CRC register only has to be combined with the CRC of the tiles affected by that drawcall. Besides this, even though multiple primitives of the same drawcall may overlap the same tile, the Constant CRC should be considered only once per tile.

Rendering Elimination uses a bitmap to solve these issues. The bitmap has a length equal to the number of tiles that the Frame Buffer is divided into. If a position of the bitmap is set, it means that the Constant CRC has already been combined into the signature for that tile. Whenever the GPU receives a new set of constants after having processed one or more drawcalls, the bitmap is cleared and the constants are signed and stored in the Constant CRC register. For all the following primitives, for every tile identifier popped from the OT Queue, the bitmap is queried to check whether that tile has already combined the signature of the constants into its signature. If so, the previous CRC of the tile is only updated with the value stored in the Primitive CRC register. Otherwise, the bit in the bitmap position corresponding to that tile is set and the previous CRC of the tile is updated twice: first with the contents of the Constants CRC, and second with the Primitive CRC, by making the Accumulate CRC unit to select the appropiate shift amount in each step.

\subsection{Compute CRC and Accumulate CRC Unit Architectures}
\graphicspath{{RenderingElimination/}}

\begin{algorithm}[ht]
\begin{algorithmic}
 \State $CRC_{Out} = 0$
 \State $ShiftAmount = 0$
 \For{64-bit subblock $A_{i}$ in submessage $A$}
 \State $CRC_{A_{i}} = ComputeCRC (A_{i})$
 \State $CRC_{Temporary} = ComputeCRC (CRC_{Out} << 64)$
 \State $CRC_{Out} = CRC_{A_{i}} \oplus CRC_{Temporary}$ 
 \State $ShiftAmount = ShiftAmount + 1$
 \EndFor
 \caption{Compute CRC Unit, Incremental \newline Computation}
 \label{alg:compute}
 \end{algorithmic}
\end{algorithm}

\begin{figure}
 \centering
 \includegraphics[width=0.40\textwidth]{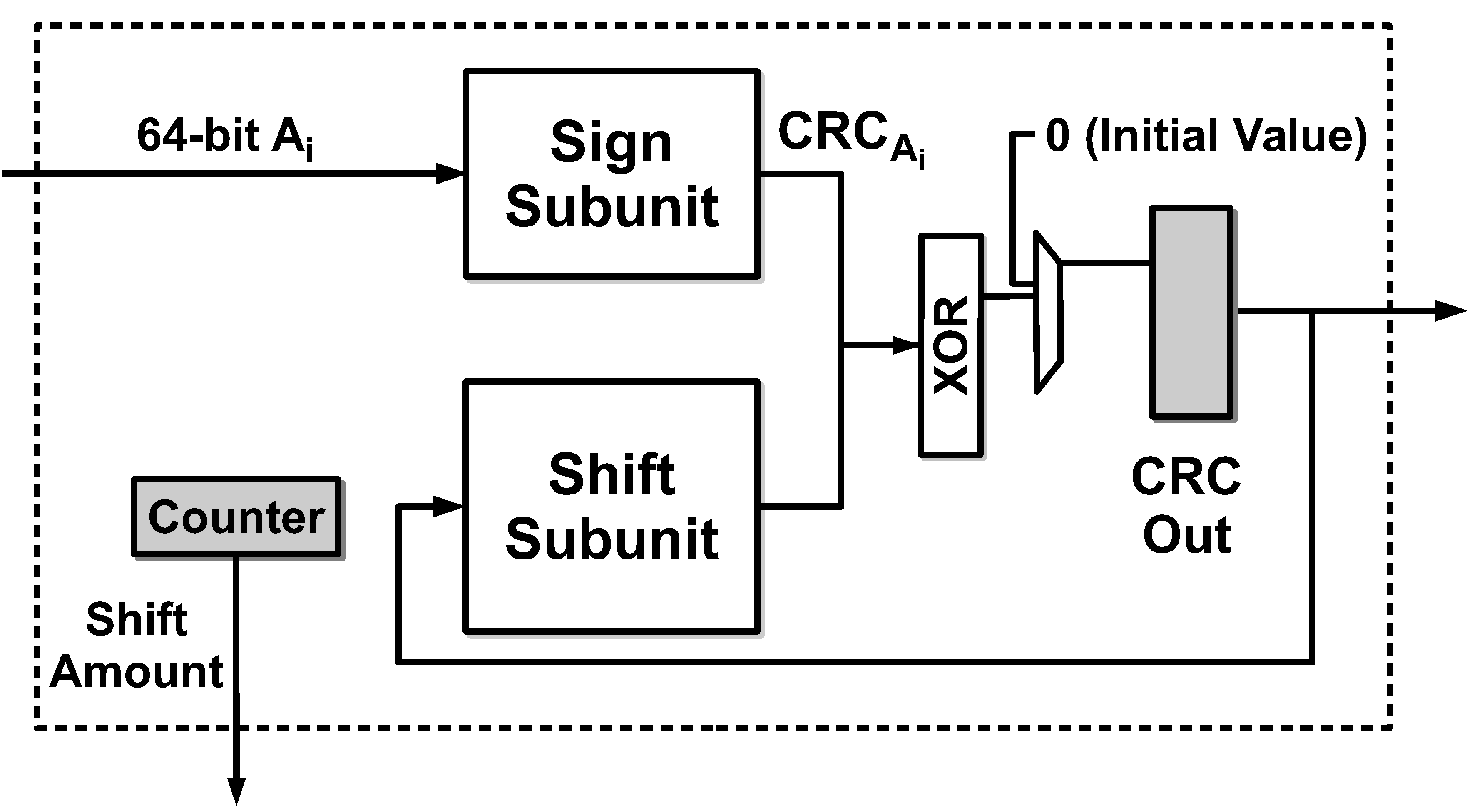}
 \caption{Compute CRC Unit block diagram.}
 \label{fig:computesubunit}
\vspace{-1em} 
\end{figure}

The \textbf{Compute CRC unit} implements the first two steps in the loop of Algorithm \ref{alg:incremental}, computing the CRC of a block consisting of a primitive or a set of constants and determining the length of the block. Since the length of such blocks is not fixed, the Compute CRC unit is architected to incrementally compute the CRC32 of a block by breaking it into subblocks of fixed length (64 bits) and recursively applying Algorithm \ref{alg:incremental}. The resulting procedure is detailed in Algorithm \ref{alg:compute}. Namely, the Compute CRC unit has a similar internal structure as the Signature Unit, as shown in Figure \ref{fig:computesubunit}. It consists of two subunits and the $CRC_{Out}$ register (initialized to zero).
The \textit{Sign} subunit computes the CRC32 of a fixed-length subblock and stores it into the $CRC_{Out}$ register after a bitwise XOR with the result of the \textit{Shift} subunit. In parallel, the Shift subunit computes the CRC32 of the message resulting by left-shifting 64 bits the contents of the $CRC_{Out}$ register. This process is repeated for each 64-bit subblock in the input data block received by the Compute CRC unit. The control logic of the Compute CRC unit counts the number of signed subblocks and communicates it to the Accumulate CRC unit using registers Shift Amount P (for Primitives) and Shift Amount C (for Constants), shown in Figure \ref{fig:architecture}.

\begin{algorithm}[ht]
\begin{algorithmic}
 \State $CRC_{Accum} = SignatureBuffer[tile]$
 \For{$k \gets 1$ \textbf{to}  $ShiftAmount$} 
 \State $CRC_{Accum} = ComputeCRC (CRC_{Accum} << 64)$
 \EndFor
 \caption{Accumulate CRC Unit, Incremental \newline Computation}
 \label{alg:accumulate}
 \end{algorithmic}
\end{algorithm}

\begin{figure}
 \centering
 \includegraphics[width=0.40\textwidth]{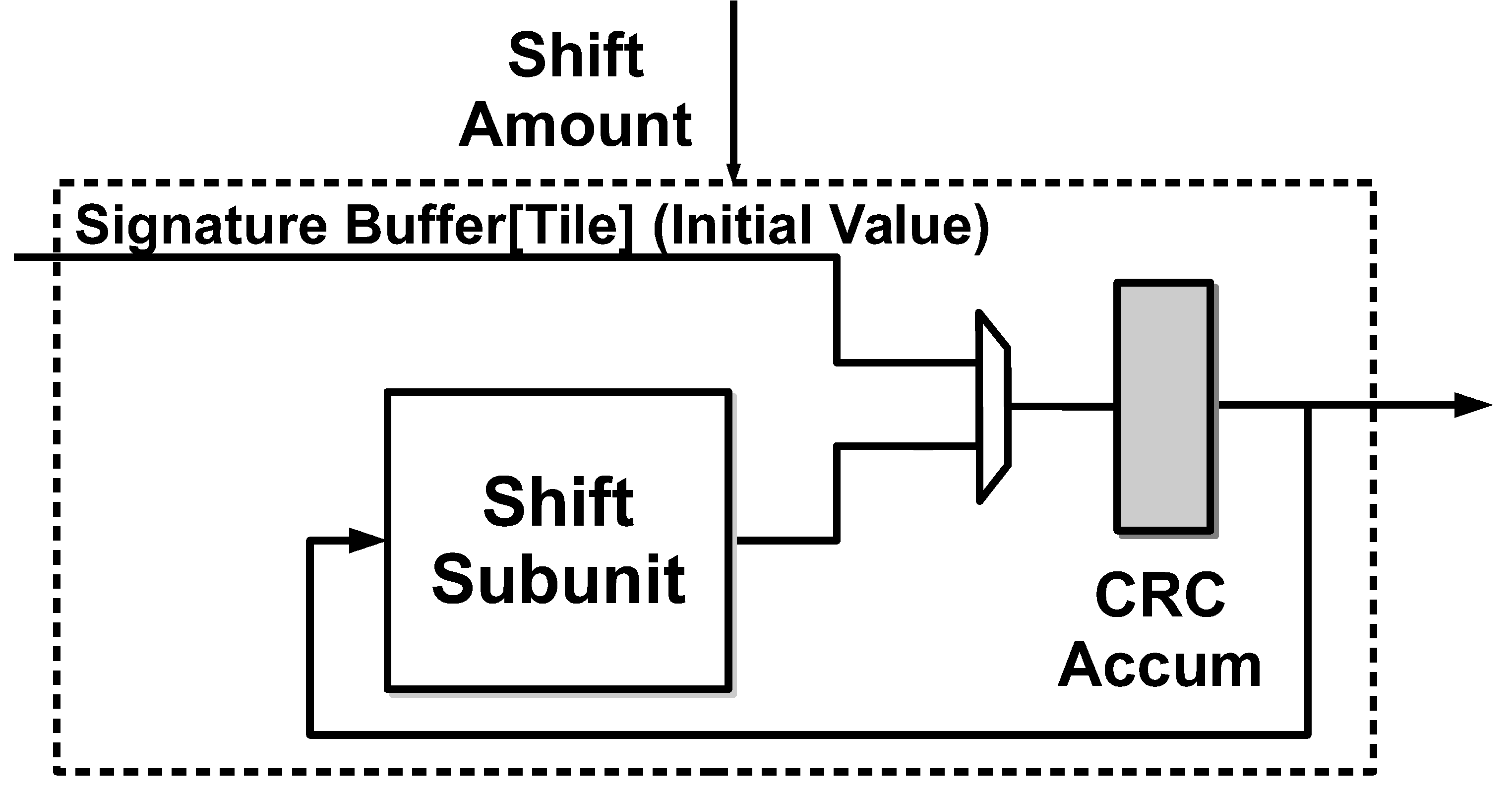}
 \caption{Accumulate CRC Unit block diagram.}
 \label{fig:accumulatesubunit}
   \vspace{-1.5em}
\end{figure}

The \textbf{Accumulate CRC unit} implements the third step in the loop of Algorithm \ref{alg:incremental}, that computes the CRC of a message consisting of the partial CRC of a tile (stored in the Signature Buffer) left-shifted by as many zeros as the length of the block to accumulate (the one fed to the Compute CRC unit). Since the length of this block is variable, it is also variable the amount to shift, hence the length of the resultant message to be signed by the Accumulate CRC unit. Therefore, this unit follows an incremental procedure to compute the CRC, as detailed in Algorithm \ref{alg:accumulate}. Note that, while the Accumulate CRC unit follows the same incremental approach as the Compute CRC unit, the accumulated blocks are always zero (they come from a left shift). Therefore, each iteration only requires to shift and re-sign the CRC32 computed on the preceding iteration and, consequently, the Accumulate CRC unit only consists of a Shift subunit, as shown in Figure \ref{fig:accumulatesubunit}.

\begin{figure}[ht]
 \centering
 \includegraphics[width=0.47\textwidth]{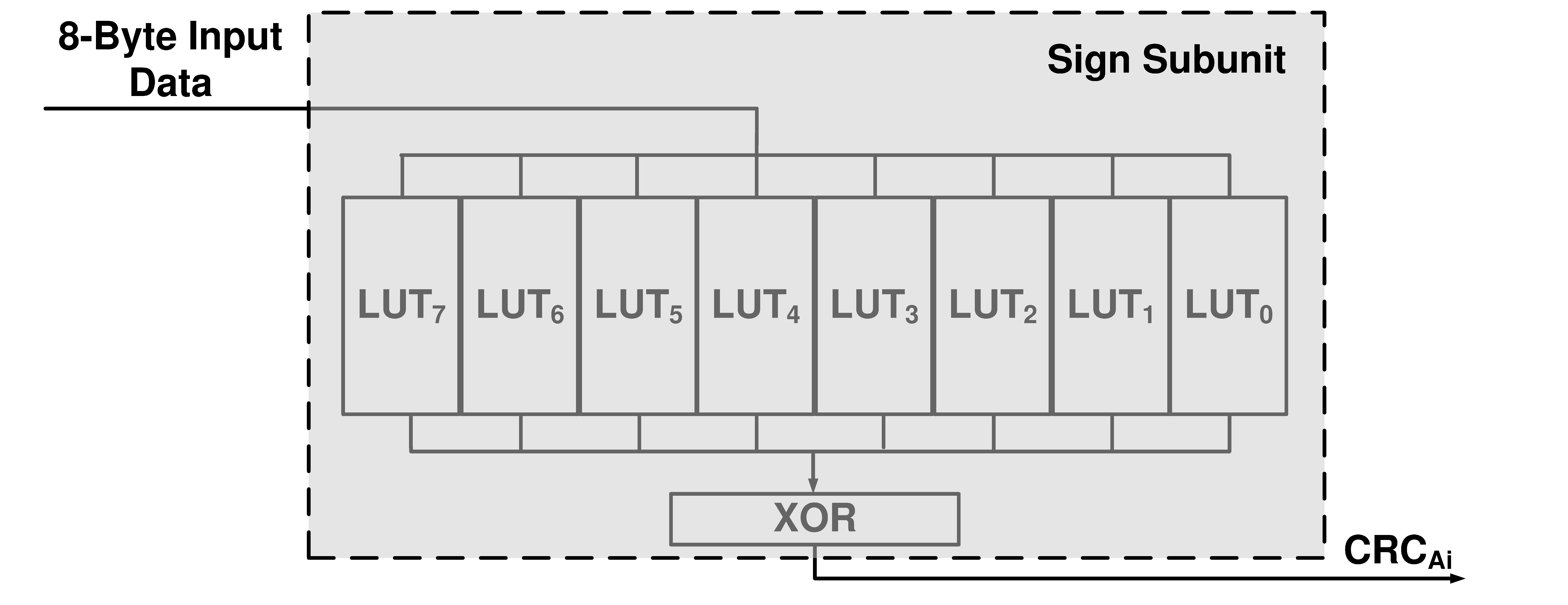}
 \caption{Architecture of the Sign subunit.}
 \label{fig:computeLUTs}
\end{figure}

Figure \ref{fig:computeLUTs} shows the \textbf{Sign subunit} architecture, which computes the CRC32 of a 64-bit subblock using the table-based approach of \cite{sun2013high}, and described in Section \ref{sec:parallel}. Each byte in the subblock is independently processed by accessing a specific LUT. The output of the Shift subunit is the bitwise XOR of the results of the 8 LUTs.

\begin{figure}[ht]
 \centering
 \includegraphics[width=0.4\textwidth]{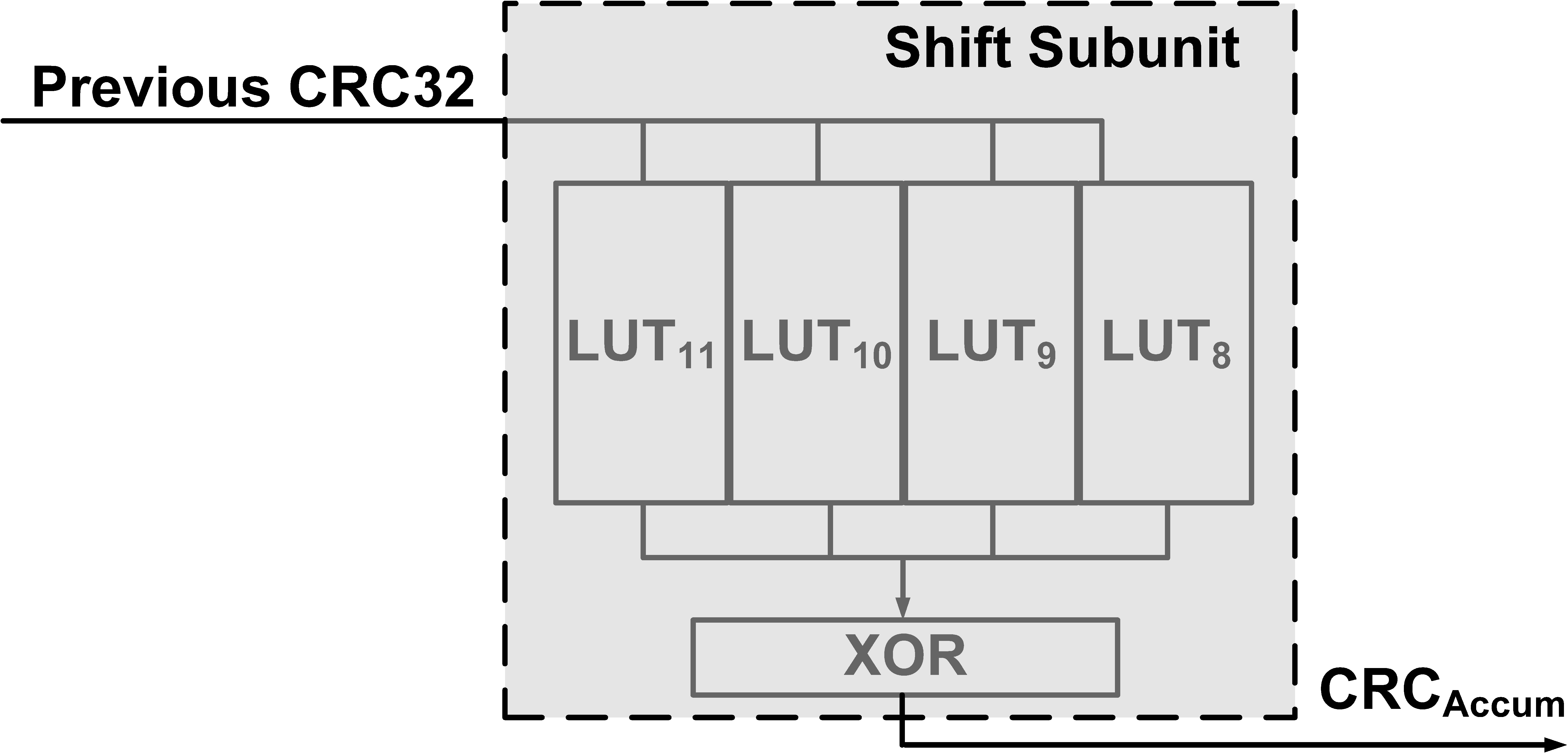}
 \caption{Architecture of the Shift subunit.}
 \label{fig:accumulateLUTs}
\end{figure}

Figure \ref{fig:accumulateLUTs} shows the \textbf{Shift subunit} architecture, which computes the CRC32 of the 64-bit message that results from a 32-bit input block shifted with 32 zeros. The design is analogous to the Sign subunit, and uses the table-based approach described in Section \ref{sec:parallel}.

The choice of the subblock size for the Compute CRC unit is determined by several tradeoffs: the length of a submessage has to be multiple of the length of the whole message, but very small submessages imply a larger number of cycles to compute the signature. Conversely, long submessages require more LUT storage, which causes energy and area overheads.

Experimentally, we have determined that subblocks of size 8 bytes signed with eight 1-KB LUTs incur in small time and energy overheads, as shown in Section \ref{sec:results}. The average command that updates constants modifies 16 values. A subblock of length 8 bytes corresponds to 2 of those values and, therefore, computing the signature for the average constant input data requires 8 cycles. Regarding primitives, the size of the data of an attribute is 48 bytes, which correspond to 3 vertices defined by four 4-byte components each. The average number of attributes per primitive is 3 and, thus, computing the signature for the average primitive requires 18 cycles.

\section{Evaluation Methodology}\label{sec:methodology}
In this section we briefly describe the simulation infrastructure and the set of benchmarks employed in the experiments to evaluate Rendering Elimination and Transaction Elimination techniques. The implementation of Transaction Elimination is also presented in this section.  

\subsection{GPU Simulation Framework}
\graphicspath{{EvaluationMethodology/}}

In order to evaluate our proposal (Rendering Elimination), as well as Transaction Elimination we employ Teapot~\cite{Arnau:2013:TTE:2464996.2464999}. Teapot is a GPU simulation framework that allows to run unmodified Android applications and evaluate performance and energy consumption of the GPU. Table~\ref{tab:simulator} shows the parameters employed in our simulations in order to model an architecture resembling an ARM Mali-450 GPU~\cite{mali450}. The Mali 400 MP series is the most deployed Mali GPU with around 19\% of the mobile GPU market~\cite{gpu_market}.

\begin{figure}[ht!]
\centering
   \includegraphics[width=\linewidth]{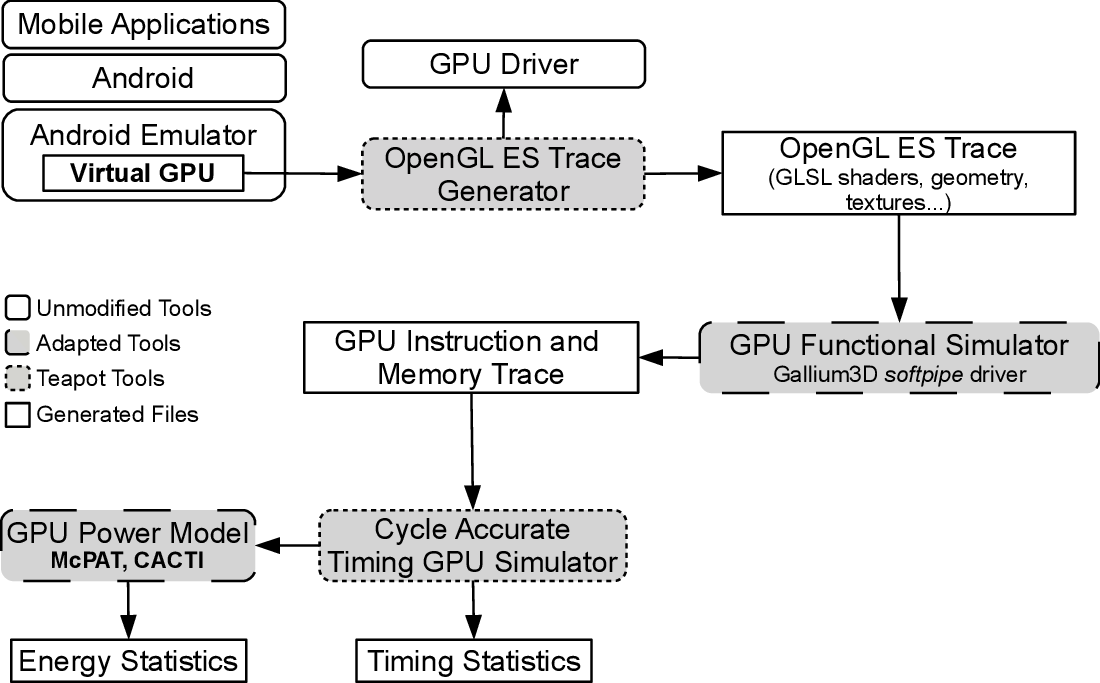}
\caption[]{Teapot simulation framework.}
\label{fig:teapot}
\end{figure}

As depicted in Figure~\ref{fig:teapot}, Teapot is comprised of three main components: OpenGL trace generator, GPU functional simulation and GPU cycle-accurate simulation. The workloads are executed in the Android Emulator deployed in the Android Studio~\cite{android_studio}. While the application is running, the OpenGL trace generator intercepts and stores all the OpenGL commands that the Android Emulator sends to the GPU driver. The OpenGL commands trace that is generated is later fed to an instrumented version of \textit{Softpipe}. \textit{Softpipe} is a software renderer included in Gallium3D, a well-known architecture for building 3D graphics drivers. Our instrumented \textit{Softpipe} executes the OpenGL commands and creates a GPU trace including information of the different stages of the graphics pipeline (memory accesses, shader instructions, vertices, primitives, fragments, texels, samplers, etc). The GPU trace is used by the cycle-accurate simulator, which gathers activity factors of all the components included in the modeled TBR architecture and reports timing as well as power consumption. Regarding the power model, McPAT~\cite{mcpat} provides energy estimations for the processors and the caches included in the GPU. We have extended McPAT using its components (SRAM, registers, XORs and MUXes, among others) to describe all the additional structures present in the architecture presented in Section \ref{sec:re}: the Signature Buffer, the CRC LUTs, the OT Queue and the constant bitmap, as well as all necessary registers and combinational logic. The main memory and the memory controller are simulated with DRAMSim2~\cite{rosenfeld2011dramsim2}. 

\begin{table}[ht!]
	\renewcommand{\arraystretch}{1.1}    
    \centering
    \caption{GPU Simulation Parameters.}
    \label{tab:simulator}
     \begin{adjustbox}{max width=\textwidth,max totalheight=0.45\textheight,keepaspectratio}
	\begin{tabular}{ l  p{5.5cm} }
	  \toprule	  
	  \multicolumn{2}{c}{\textbf{Baseline GPU Parameters}} \\
	  \bottomrule	  	  
	  Tech Specs & 400 MHz, 1 V, 32 nm \\
	  Screen Resolution & 1196x768 \\
	  Tile Size & 16x16 pixels \\
      \toprule
	  \multicolumn{2}{c}{\textbf{Main Memory}} \\
      \bottomrule
	  Latency &	50-100 cycles \\
	  Bandwidth	&	4 bytes/cycle (dual channel LPDDR3) \\
	  Size	&	1 GB  \\
      \toprule
	  \multicolumn{2}{c}{\textbf{Queues}} \\
      \bottomrule  
	  Vertex (2x)	 &	16 entries, 136 bytes/entry \\
	  Triangle, Tile &	16 entries, 388 bytes/entry \\
	  Fragment	 &	64 entries, 233 bytes/entry \\
      \toprule
	  \multicolumn{2}{c}{\textbf{Caches}} \\
      \bottomrule	  
	  Vertex Cache	&	64 bytes/line, 2-way, 4 KB, 1 bank, 1 cycle \\
	  Texture Caches (4x)	&	64 bytes/line, 2-way, 8 KB, 1 bank, 1 cycle \\
	  Tile Cache	&	64 bytes/line, 8-way, 128 KB, 8 banks, 1 cycle \\
	  L2 Cache	& 64 bytes/line, 8-way, 256 KB, 8 banks, 2 cycles \\
	  Color Buffer & 64 bytes/line, 1-way, 1 KB, 1 bank, 1 cycle \\
	  Depth Buffer & 64 bytes/line, 1-way, 1 KB, 1 bank, 1 cycle \\
      \toprule
	  \multicolumn{2}{c}{\textbf{Non-programmable stages}} \\
      \bottomrule  
	  Primitive assembly	&	1 triangle/cycle \\
	  Rasterizer 	&	16 attributes/cycle \\
	  Early Z test	&	32 in-flight quad-fragments, 1 Depth Buffer\\
      \toprule
	  \multicolumn{2}{c}{\textbf{Programmable stages}} \\
      \bottomrule
	  Vertex Processor	&	1 vertex processor \\
	  Fragment Processor	&	4 fragment processors \\
	  \bottomrule
	\end{tabular}
	\end{adjustbox}
\vspace{-1em}
\end{table}

\subsection{Benchmark Suite}
Table \ref{tab:benchmarks} shows the set of benchmarks analyzed to evaluate our technique, which consists of ten commercial Android graphics applications. Our set of benchmarks includes both 2D and 3D games, applications that stress the GPU further than other commonly used applications in battery-operated devices. Among the 3D games we include workloads with simple 3D models such as \textit{Tigerball}, and workloads with more sophisticated 3D models and scenes such as \textit{Modern Strike} and \textit{Temple Run}. The workloads included in our set of benchmarks are representative of the current landscape of smartphone games ecosystem as it includes popular Android games for smartphones and tablets. These applications have millions of downloads according to Google Play~\cite{googleplay}, some of them surpassing 500 million downloads. 

\begin{table}[ht]
\centering
\caption{Benchmark suite.}
\label{tab:benchmarks}
\begin{tabular}{@{}lllc@{}}
\toprule
\textbf{Benchmark}      & \textbf{Alias} & \textbf{Genre} & \textbf{Type} \\
\midrule
Angry Birds             & abi            & Arcade               & 2D\\
Candy Crush Saga        & ccs            & Puzzle               & 2D\\
Castle Defense          & cde             & Tower Defense        & 2D\\
Clash of Clans          & coc            & MMO Strategy         & 3D\\
Crazy Snowboard         & csn             & Arcade               & 3D\\
Cut the Rope            & ctr            & Puzzle               & 2D\\
Hopeless                & hop            & Survival Horror      & 2D\\
Modern Strike           & mst             & First Person Shooter & 3D\\
Temple Run              & ter             & Platform             & 3D\\
Tigerball               & tib            & Physics Puzzle       & 3D\\
\bottomrule
\end{tabular}
\vspace{-2em}
\end{table}

\subsection{Transaction Elimination}
\begin{figure}[ht]
\centering
   \includegraphics[width=0.95\linewidth]{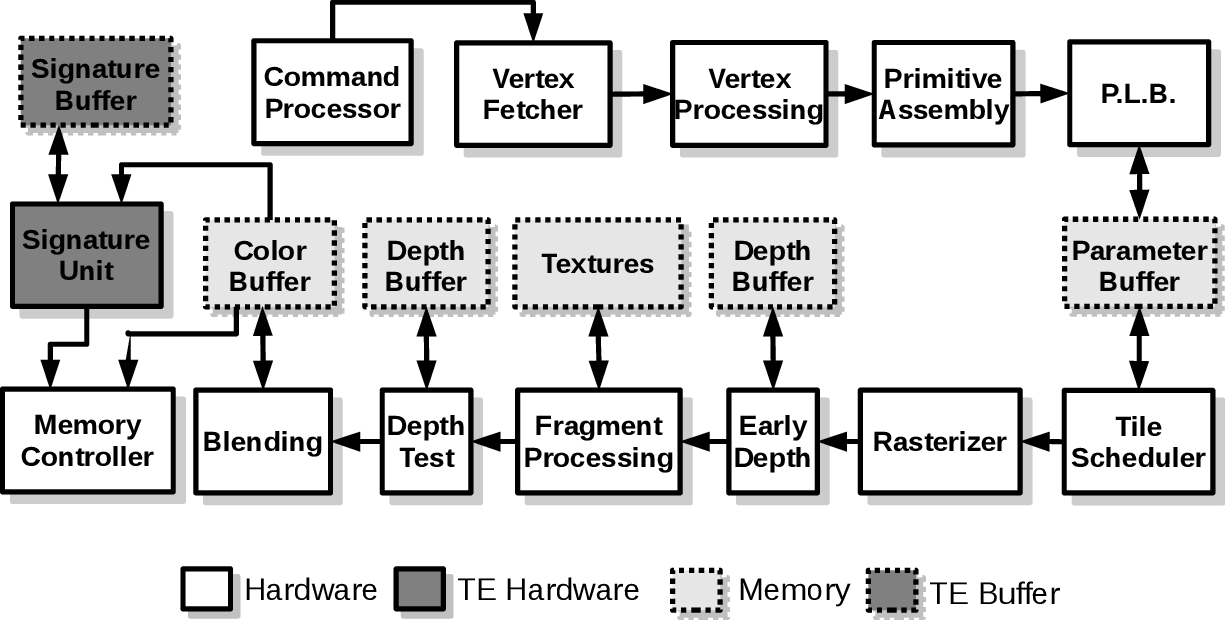}
   \caption{Graphics Pipeline including Transaction Elimination.}
   \label{fig:TE_b} 
  \vspace{-0.5em}   
\end{figure}

\graphicspath{{EvaluationMethodology/}}
Transaction Elimination (TE)~\cite{TE} is a technique that reduces   main memory bandwidth by avoiding the flush of the Color Buffer in tiles that have the same color as in the preceding frame. Since the reuse distance of two tiles is an entire frame, tile equality is not performed by comparing the colors of all the pixels of a tile but rather signatures of those colors. Whenever a tile has finished being rendered, its colors (the contents in the Color Buffer) are hashed into a signature and compared to the signature of the same tile for the previous frame. If the two signatures are equal, the newly generated colors are not written into the Frame Buffer. Although the exact details of this technique in commercial systems are not fully disclosed, we have modified our cycle-accurate simulator to model an efficient implementation and compare it with our proposed approach. Figure \ref{fig:TE_b} presents the extra hardware added in the pipeline to perform Transaction Elimination.

In our TE evaluation, we consider the energy overheads caused by the Signature Buffer and the Compute CRC unit, but do not add any execution time overhead: while we count the number of accesses to the Compute CRC unit to report energy, we ideally assume that the signature for a Color Buffer does not require any execution cycles.

For both the evaluation of Rendering Elimination and Transaction Elimination, we consider the common case in current GPUs in which the memory system has not only one but two Frame Buffers. This allows the display to read from one (called Front Buffer) while the GPU processes the following frame by writing into a different memory region (Back Buffer) without causing visual artifacts. The Front and Back buffers are periodically swapped so that the display presents new frames at the appropriate frame rate. With this approach, tiles have to be compared not with the frame being displayed but with one prior, since the potential transactions to eliminate occur between the GPU and the Back Buffer. The Signature Buffer, therefore, contains signatures spanning two frames: the set generated when the GPU processes a frame and writes into the Back Buffer and the set for the Front Buffer, that will be used to compare tiles when the buffers are swapped.
 \label{sec:te}

\section{Experimental Results}\label{sec:results}
\graphicspath{{Results/}}

\begin{figure*}[ht!]
\centering
\subfloat[]{\includegraphics[trim = 0mm 0mm -5mm -10mm, clip, width=0.92\columnwidth]{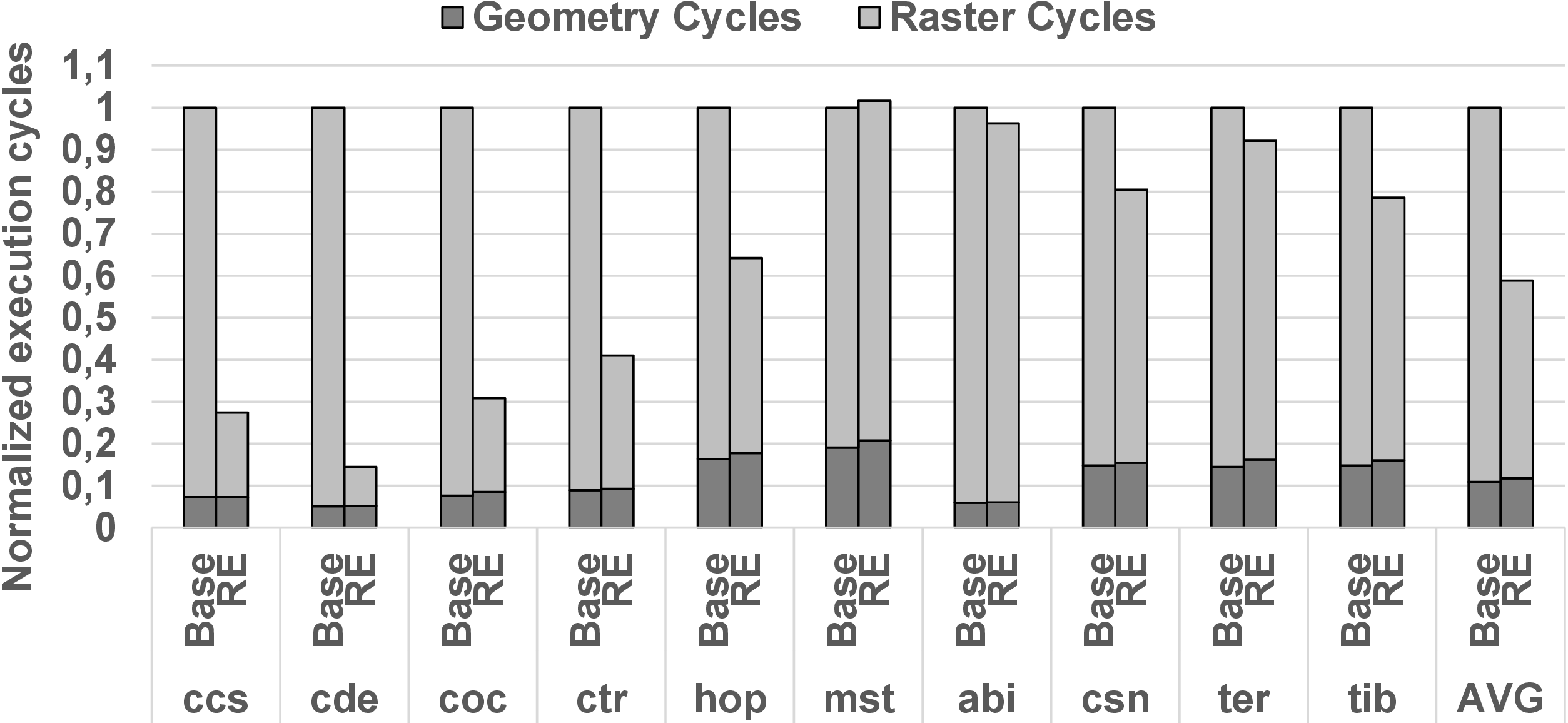}\label{fig:cycles_results}}
\subfloat[]{\includegraphics[trim = 0mm 0mm 0mm -10mm, clip, width=0.92\columnwidth]{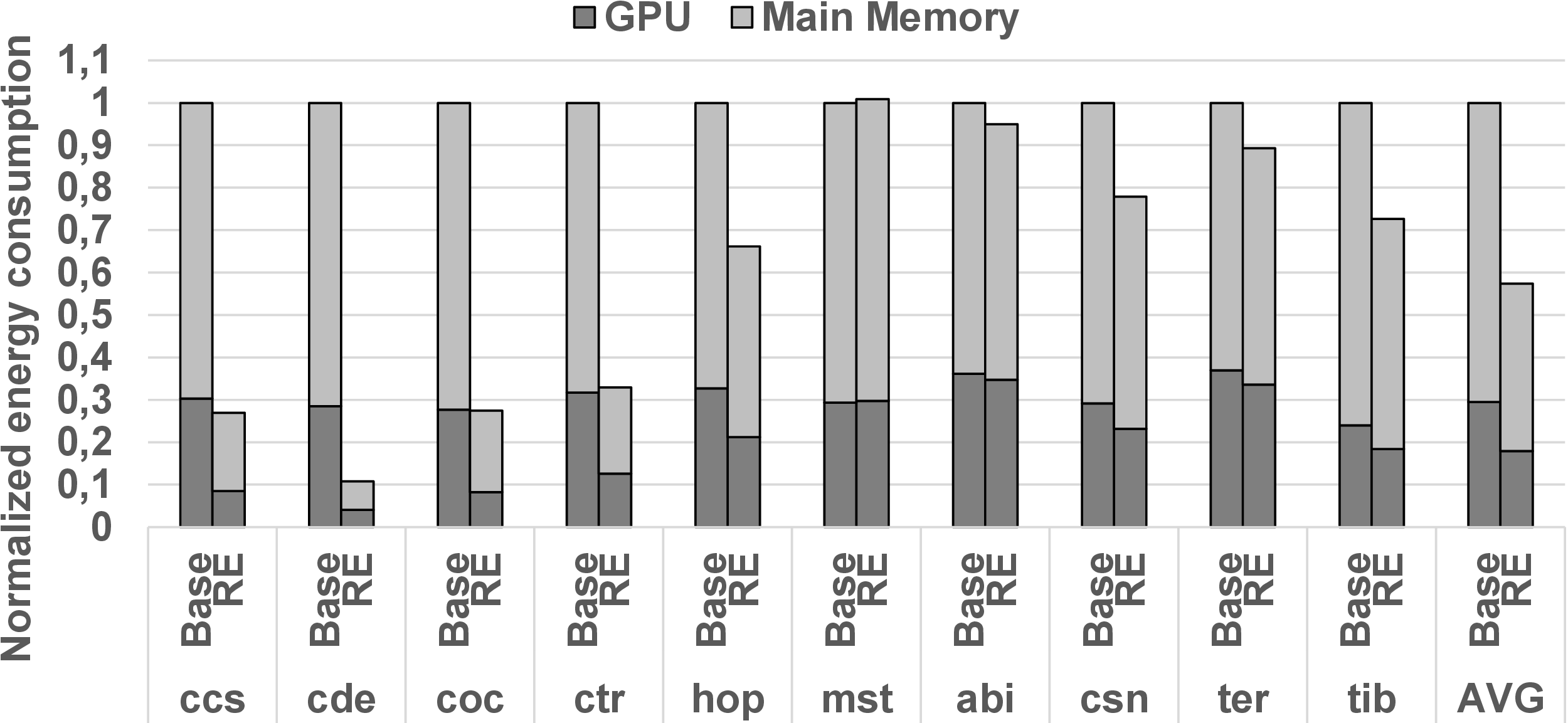}\label{fig:energy_results}}
\caption{Rendering Elimination compared against Baseline GPU: \protect\subref{fig:cycles_results} Execution cycles. \protect\subref{fig:energy_results} Energy consumption.}
\label{fig:REresults}
 \vspace{-2em}
\end{figure*}

\begin{figure*}[ht!]
\centering
\subfloat[]{\includegraphics[trim = 0mm 0mm -5mm -10mm, clip,width=0.92\columnwidth]{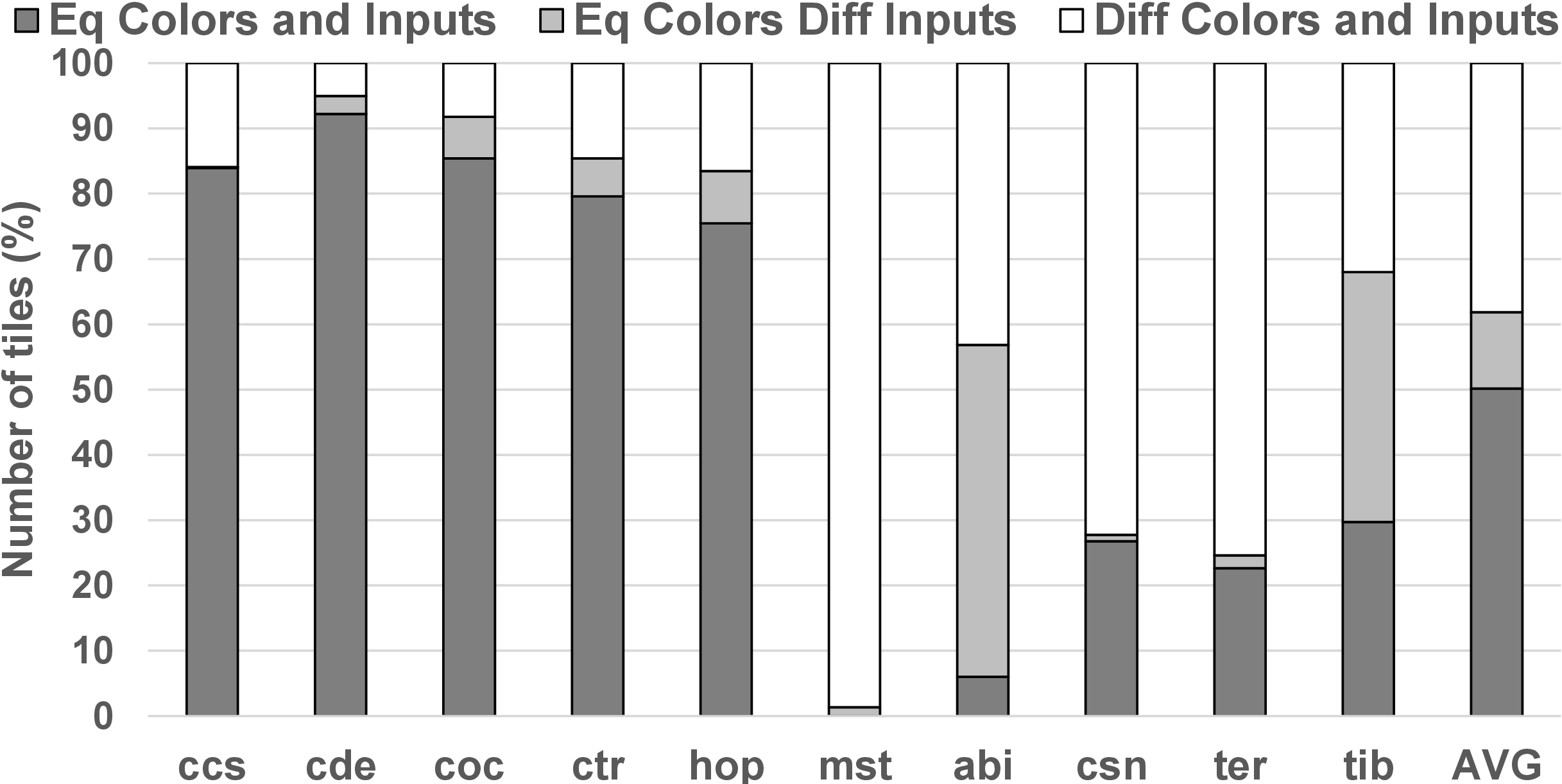}\label{fig:equaltiles_r}}
\subfloat[]{\includegraphics[trim = 0mm 0mm 0mm -10mm, clip,width=0.92\columnwidth]{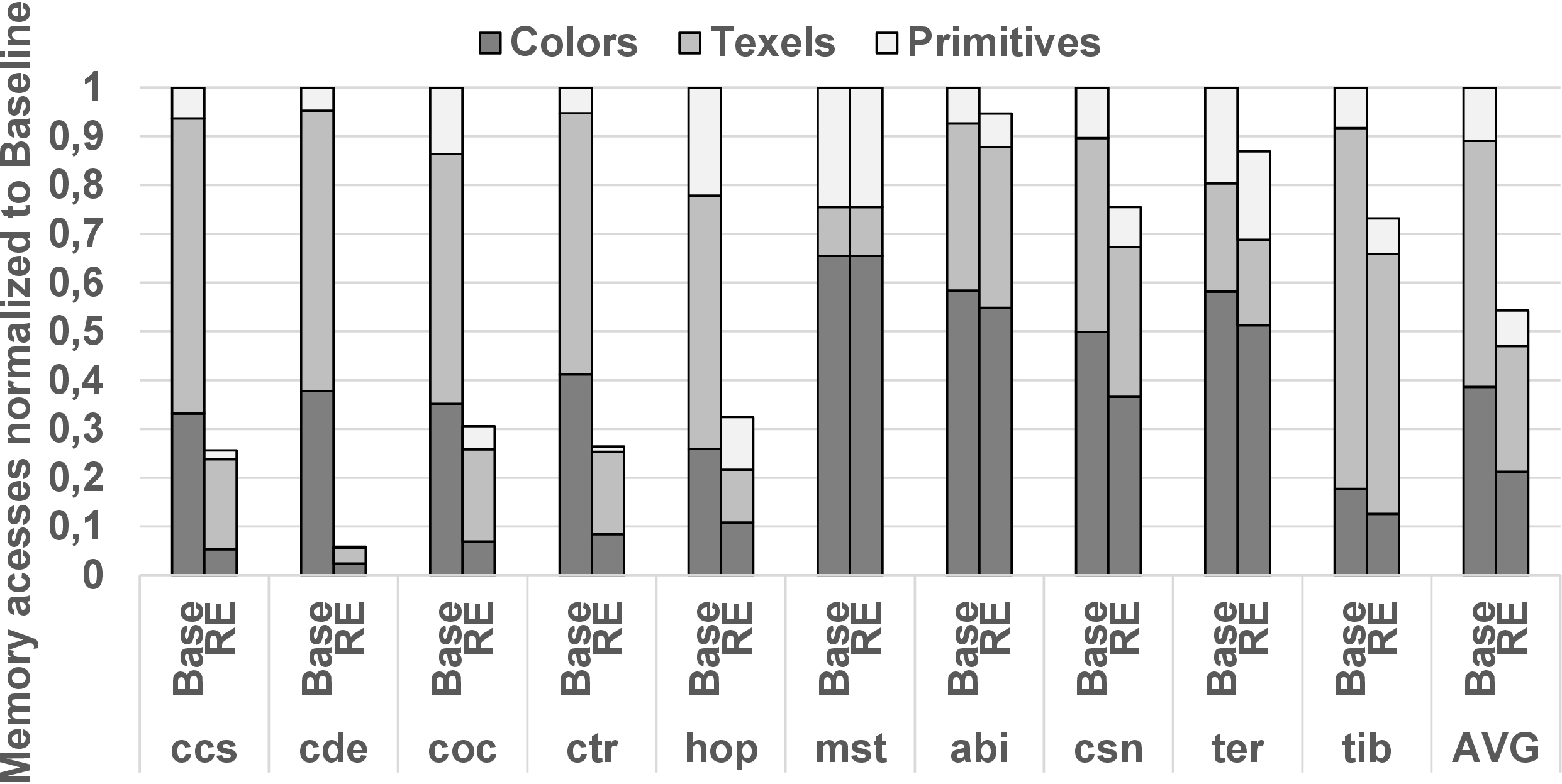}\label{fig:memory_results}}
\caption{Sources of execution time and energy reduction \protect\subref{fig:equaltiles_r} Percentage of tiles with equal color and equal inputs, equal color and different inputs, and different color and different inputs across neighboring frames. \protect\subref{fig:memory_results} RE main memory bandwidth compared to baseline GPU including Primitive reads from the Parameter Buffer, Texel fetches and Color Buffer flushes.}
\vspace{-1em}
\end{figure*}

In this Section we present the main results of Rendering Elimination over the baseline. For comparison purposes, we also evaluate Fragment Memoization \cite{Arnau:2014:ERF:2665671.2665748} and Transaction Elimination (see implementation in Section~\ref{sec:te}).

Figure~\ref{fig:cycles_results} shows execution cycles of RE for our set of benchmarks. The total cycles are normalized to those of the Baseline and divided into cycles corresponding to Geometry and Raster Pipelines. RE achieves an average execution time reduction of 42\% (1.74x speedup), yielding reductions of up to 86\% (\textit{cde}). The execution of the Raster Pipeline using RE is 2x faster than the Baseline GPU on average, with maximums of more than 10x. On the other hand, the overheads introduced by the technique are almost negligible, since the signature computation is usually overlapped by previous Geometry Pipeline stages. The pipeline is only stalled when computing signatures for primitives that cover a large amount of tiles, resulting in an overflow of the Overlapped Tiles Queue. These kind of primitives are rare, as can be seen  by the fact that, on average, only a 0.64\% additional geometry cycles are introduced. The overhead of comparing the signatures is even smaller. Considering that accessing the corresponding Signature Buffer entry and performing a simple comparison takes a few cycles while skipping the entire Raster Pipeline can save thousands, these tiny overheads are more than offset by the large performance gains. Such overheads only result in performance loss in benchmarks that lack redundant tiles and cannot leverage RE at all. Even in those cases, the performance impact is smaller than 1\%, as it can be seen in \textit{mst}.

Figure \ref{fig:energy_results} shows the GPU energy consumption (considering both static and dynamic) when using RE for our set of benchmarks, normalized to the baseline. The total energy is split into two parts: energy spent by the GPU in accessing main memory and energy spent in other activities. As shown, RE brings about an average 43\% reduction of the energy consumed by the system, with a 38\% reduction of the energy consumed by the GPU and 48\% reduction of the energy consumed by main memory. Moreover, RE provides enormous energy savings for benchmarks such as \textit{ccs} or \textit{cde}, reducing 90\% of the overall energy consumed by the baseline. In \textit{mst}, a benchmark that does not take advantage of RE, the energy overheads are smaller than 1\%. Regarding area, McPAT reports that the cost of the hardware added (CRC LUTs, Signature Buffer, Overlapped Tiles Queue and bitmap) incurs in less than 1\% area overhead.

These reductions in execution time and energy consumption are due to an important number of tiles bypassing the execution of the Raster Pipeline and avoiding their corresponding main memory accesses. Figure~\ref{fig:equaltiles_r} shows the average percentage of tiles that, across neighboring frames, produce the same color (the sum of bottom and mid bars) and the average percentage of tiles that change colors (top bar). The bottom bar depicts the percentage of tiles that Rendering Elimination avoids rendering, which is, on average 50\% of the tiles of a frame and 81\% of the total redundant tiles. The mid bar shows the percentage of tiles that despite having different inputs end up with the same color (12\%). The top bar presents the percentage of tiles with different inputs and different colors (38\%). Note that there is not a single occurrence of a tile that changes the color while maintaining the same inputs. Furthermore, Figure \ref{fig:equaltiles_r} reveals three different behaviors for the benchmarks analyzed depending on camera movements. The first category, (\textit{ccs} to \textit{hop}) is composed of workloads with mainly static cameras, so their scenes contain lots of redundant tiles. The second category (\textit{mst}) is composed of workloads with highly dynamic camera movements and almost no redundant tiles. The third category (\textit{abi} to \textit{tib}) behaves like the first set in some phases and like the second set in others. It can be seen that there is a strong correlation between the number of detected redundant tiles presented on Figure~\ref{fig:equaltiles_r} and the speedup and energy savings reported in Figure \ref{fig:REresults}.

Eliminating redundant tiles not only reduces the activity of the GPU but it also eliminates all the associated memory accesses. Figure \ref{fig:memory_results} plots the amount of main memory traffic generated by the Raster Pipeline, normalized to the baseline. The total traffic is split into three parts: accesses generated by the Tile Cache when reading primitives from the Parameter Buffer, accesses generated by the Texture Cache when fetching textures in the fragment shaders and accesses generated by flushing the on-chip Color Buffer to the Frame Buffer. As it is shown, RE achieves a significant drop in traffic to main memory (48\% on average).

\subsection{RE vs Fragment Memoization and Transaction Elimination}\label{sec:comparison}
\graphicspath{{Results/}}

\begin{figure}[ht]
\centering
   \includegraphics[width=0.92\linewidth]{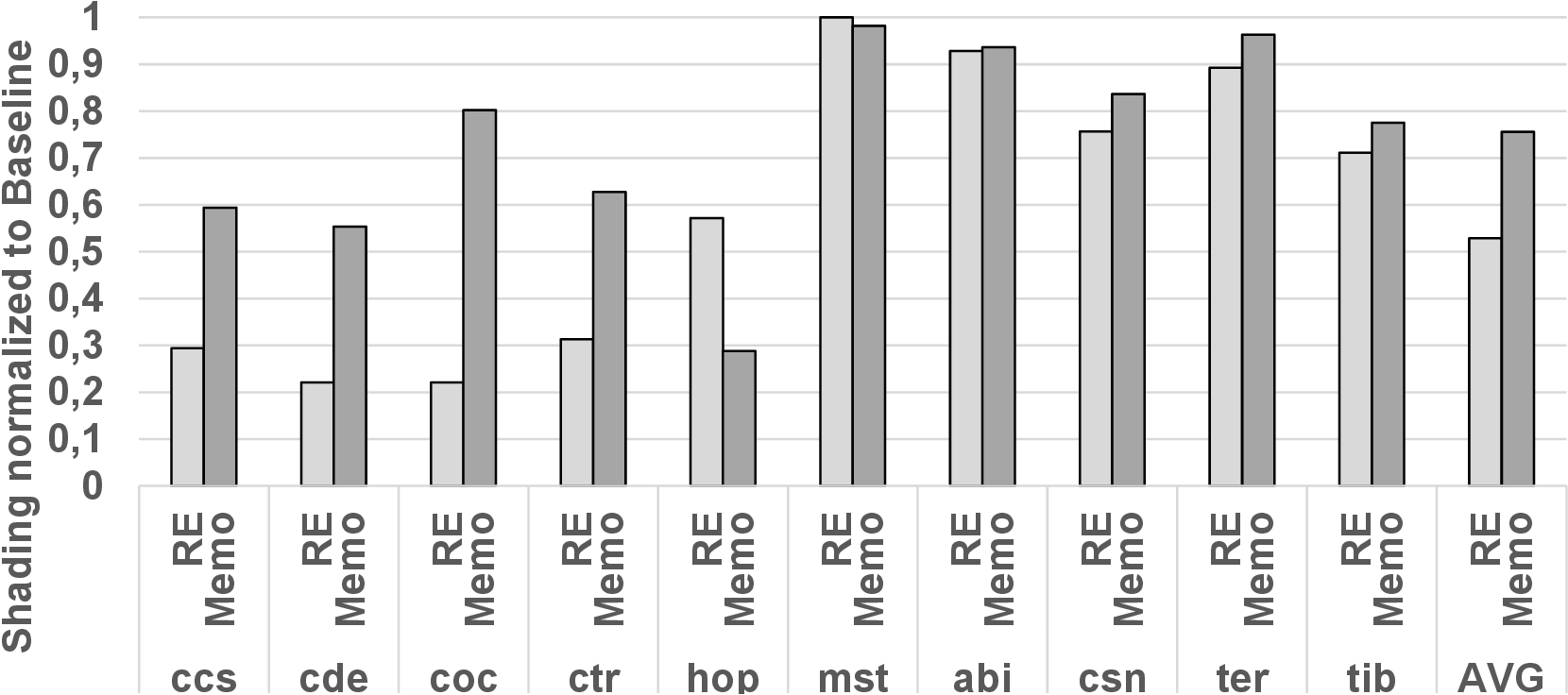}
\caption[]{Fragments shaded with RE and PFR-aided Fragment Memoization normalized to baseline.}
 \label{fig:fragments_TE_RE} 
  \vspace{-0.5em}
\end{figure}

Figure \ref{fig:fragments_TE_RE} compares the number of fragments shaded by RE to those shaded by the technique proposed by Arnau et al. \cite{Arnau:2014:ERF:2665671.2665748}, which performs fragment  memoization but requires rendering multiple frames in parallel. Note also that our approach is able to skip more pipeline stages and their corresponding main memory accesses (see Figure \ref{fig:f2fraster}). We run an experiment to compare the amount of reused fragments by each technique. We modelled Fragment Memoization as originally proposed, with 2-frames in parallel and a 32-bit hash that discards the screen coordinates, but we augmented their default 512-entry 4-way LUT to 2048 entries to better compare to the chip area of RE. As shown, RE reuses much more fragments in the majority of benchmarks. One would expect that, by working at a fragment granularity, memoization could discover more redundancy than working at a tile level. However such granularity also requires a bigger storage and, as already pointed out in their paper, a realistic space-limited LUT only captures on average 60\% of that potential, whereas RE captures all of the redundant tiles with equal inputs. The only notable exception is \textit{hop}, because it renders a large portion of the screen with a small number of repeated fragments, most of them completely black, thus heavily reducing the pressure on the LUT storage, but this is a rather rare case. Moreover, because of the large reuse distance between redundant fragments, Fragment Memoization requires significant modifications in the pipeline to enable rendering of multiple frames in parallel. While executing two frames in parallel has benefits beyond memoization, it has two major drawbacks that RE does not. First, it implies a significant re-design of the whole GPU. Second, it generates input response lag because of the parallel frame rendering process. To alleviate this side effect it must be disabled during frames where the user introduce inputs.

\begin{figure*}[ht!]
\centering
\subfloat[]{\includegraphics[trim = 0mm 0mm -5mm -10mm, clip,width=0.96\columnwidth]{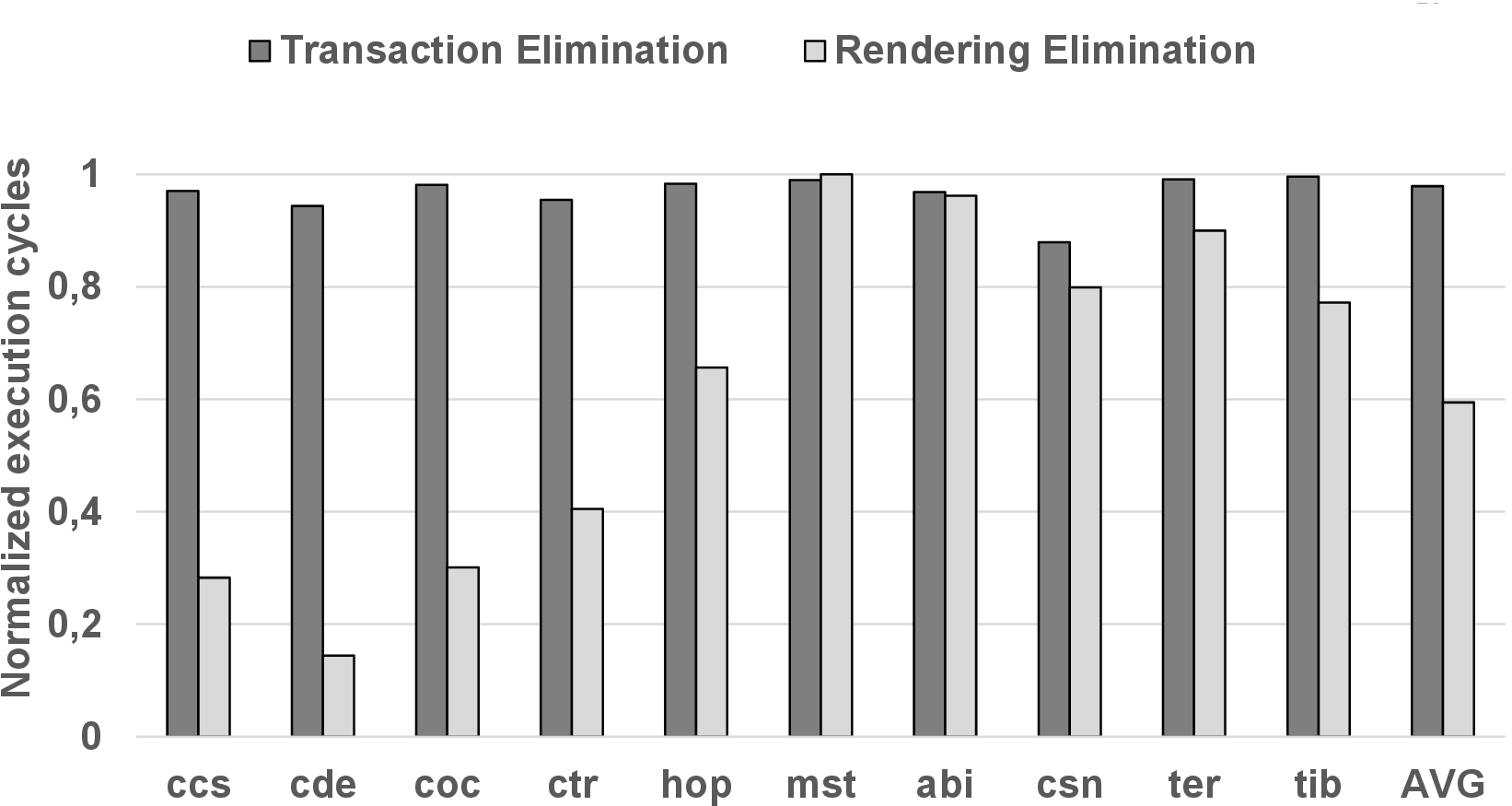}\label{fig:re_te_speedup}}
\subfloat[]{\includegraphics[trim = 0mm 0mm 0mm -10mm, clip,width=0.96\columnwidth]{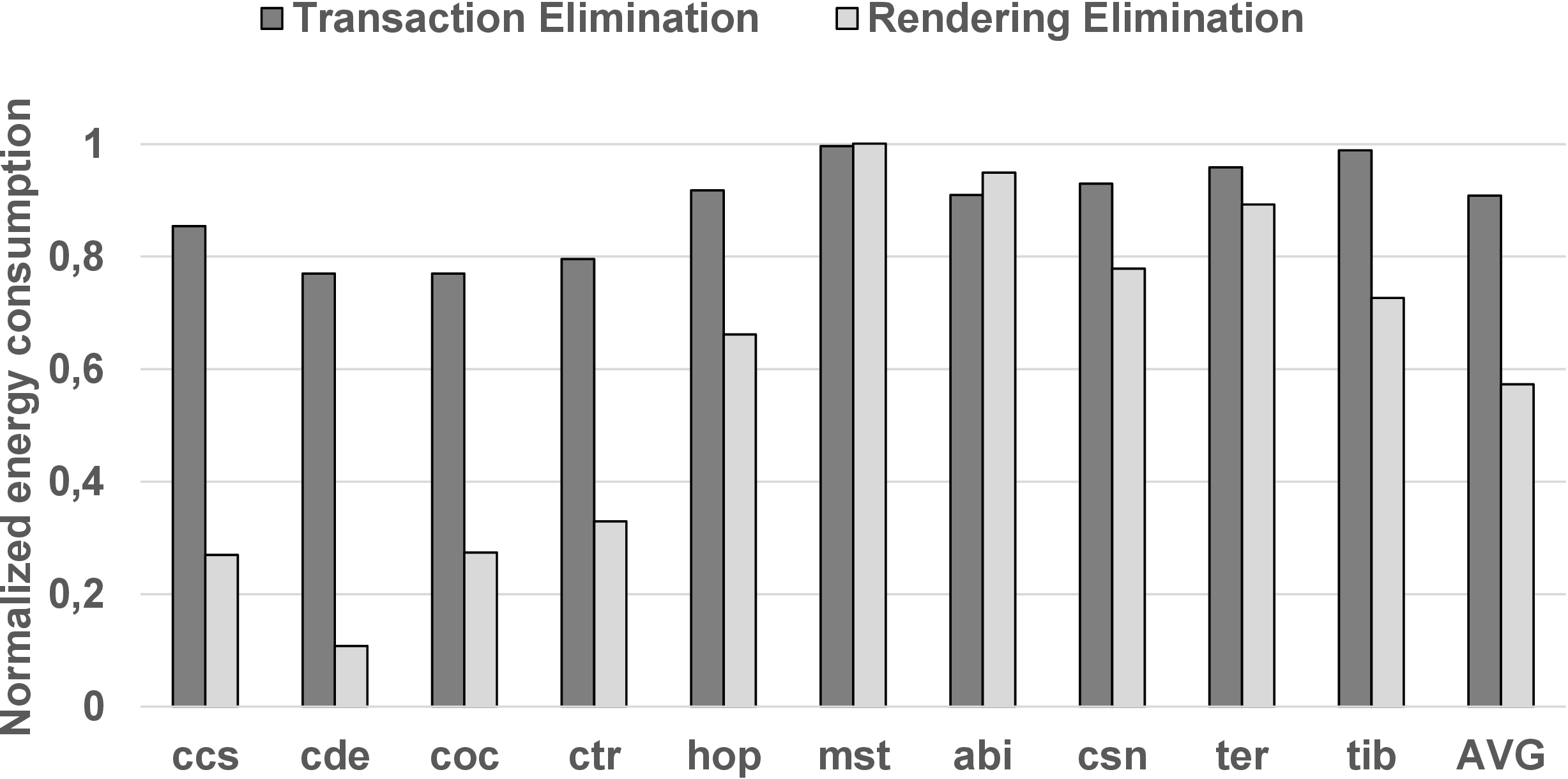}\label{fig:re_te_energy}}
\caption{Comparison of RE and Transaction Elimination against Baseline GPU. \protect\subref{fig:re_te_speedup} Execution cycles.  \protect\subref{fig:re_te_energy} Energy consumption.}
\label{fig:re_te}
  \vspace{-1.5em}
\end{figure*}

Figure \ref{fig:re_te} compares the benefits of Rendering Elimination over Transaction Elimination (see implementation details in Section \ref{sec:te}). Transaction Elimination (TE) avoids only the Color Buffer flushes to main memory, while RE bypasses the whole Raster Pipeline execution for redundant tiles. Therefore, while TE reduces a 9\% the energy consumption with respect to the baseline GPU, RE outperforms it and achieves a reduction of 43\%. Note that in benchmarks with a large percentage of redundant tiles such as \textit{cde}, RE achieves an additional 65\% energy savings compared with TE. Moreover, since the flush of the Color Buffer represents a relatively small portion of the total time of the Raster Pipeline, RE far surpasses the performance benefits of TE.

In some cases, TE may obtain energy savings for benchmarks in which RE cannot. As Figure \ref{fig:equaltiles_r} presents, there is a subset of tiles whose rendering outputs the same color as in the preceding frame but do not have the same inputs as in the preceding frame (depicted in the mid bar). On average, this occurs for 12\% of the tiles. This phenomenon may occur, for instance, when the only differences between the two tiles happen on occluded fragments that are eventually culled by the z-test and do not contribute to the final color of the tile, or for scenes with quick camera panning movements where most of the background texture contains a single plain color. Consequently, in benchmarks where RE detects a small percentage of equal tiles, such as \textit{abi}, TE may obtain a slightly better energy savings than RE.

We refer to the above event, where the signature of two tile inputs does not match but the final color of their pixels remains unchanged, as \textit{false negatives}. False negatives do not generate errors, but reveal a broader potential for tile reuse that RE is not capable to detect. On the other hand, since tile inputs are compared using the result of a hash function, there exists the possibility of collisions or \textit{false positives}: pairs of different tile inputs that are mapped to the same signature. A false positive means that the GPU incorrectly reuses a tile that has actually changed in the current frame. However, the probability of such an event with a CRC32 signature is roughly one every 4 billion tiles, i.e., less than one tile per million frames (more than 4 hours playing). Moreover, it would be extremely difficult, or impossible, to spot the incorrect tile by a human, since it would last for only a single frame (less than 20 ms), and it would probably appear very similar to the correct tile due to frame coherency. Actually, we found zero false positives in our experiments with CRC32.




\section{Related Work}\label{sec:rw}


Hardware memoization has been widely researched to accelerate general-purpose computing by detecting blocks of instructions that repeatedly produce the same value and caching them in limited-space LUTs~\cite{lipasti1996value,sodani1997dynamic,citron1998accelerating,huang1999exploiting}. 

Several works exploit frame coherence in order to avoid the processing of redundant fragments and their corresponding main memory accesses for GPUs. Ragan-Kelley et al.~\cite{Ragan-Kelley:2011:DSG:1966394.1966396} decouples shading from visibility and employs a hardware memoization scheme that caches shading results. Arnau et al.~\cite{Arnau:2013:PFR:2523721.2523736} also propose a hardware memoization scheme to reduce redundant fragment shading that is implemented on top of a Parallel Frame Rendering pipeline to improve reuse distance. A comparison with this technique is provided in Section~\ref{sec:comparison}. RE is similar to memoization in that it remembers the signatures of previous inputs to detect redundancy. However, since RE works at a coarser granularity, instead of caching just a fraction of these signatures, it stores all of them. Furthermore, the outputs do not need to be cached, since they are already present in the Frame Buffer.

Transaction Elimination (TE)~\cite{TE} is a bandwidth saving feature included in the ARM Mali GPU that detects identical tiles between the current frame being rendered and a previously rendered frame. TE computes a CRC signature per tile. If a tile of the current frame has the same CRC as the same tile of the preceding frame the tile is redundant and it is not flushed to main memory, which produces significant energy savings. On the other hand, RE not only avoids the flush of redundant tiles to main memory, but also the execution of the entire Raster Pipeline.

Some works aim to reduce fragment shading by means of reducing the number of occluded fragments whose color is computed. Occlussion queries~\cite{10.1111:j.1467-8659.2004.00793.x, sekulic2004efficient} rasterize and test the visibility of Bounding Volumes of the objects to cull the geometry at draw command level granularity. However, the queries need to be sorted in a front-to-back order to perform well, which sets an important limitation. Other works aim to avoid fragment shading for hidden surfaces at fragment level granularity~\cite{Haines:1996:FLM:643323.643324, Clarberg:2013:SDS:2461912.2462022}. These methods propose to perform a hidden surface removal phase where geometry is rasterized and depth tested in order to identify the visible geometry that will be later fragment shaded. Unlike these works, RE does not need to perform extra rendering passes to reduce overshading. Furthermore, in some workloads RE avoids the execution of the entire rasterization pipeline for the vast majority of the frame (even for visible surfaces).

\section{Conclusions}\label{sec:conclusions}
In this paper we have presented Rendering Elimination (RE), a novel micro-architectural technique for Tile-Based Rendering GPUs that effectively reduces shading computations and memory accesses by means of culling redundant tiles across consecutive frames. Since RE detects a redundant tile before it is dispatched to the Raster Pipeline, the entire computation (which includes rasterization, depth test, fragment processing, blending, etc.) is avoided, as well as all the associated energy-consuming memory accesses to the Parameter Buffer, Textures and Frame Buffer. 

Our results show that RE outperforms state-of-the-art techniques such as Transaction Elimination or Fragment Memoization, which are only able to bypass a single pipeline stage. Compared to the baseline GPU, RE achieves an average speedup of 1.74x and reduces the GPU and main memory energy consumption by 38\% and 48\%, respectively. The hardware overhead of RE is minimum, requiring less than 1\% of the total area of the GPU, while its latency is hidden by other processes of the graphics pipeline. In terms of energy, RE incurs a negligible overhead of less than 0.5\% of the total GPU energy. Note that it is especially efficient in benchmarks with small camera movements, with speedups as high as 6.9x and energy savings up to 90\%. Nonetheless, in benchmarks without any significant amount of redundant tiles the performance impact is well smaller than 1\%.

\bibliographystyle{IEEEtran.bst}

\end{document}